\documentclass[aps,prb,reprint,groupedaddress, showpacs]{revtex4-1}
\usepackage{graphicx,graphics}
\usepackage[table]{xcolor}
\usepackage{epstopdf}
\usepackage{amsmath,amssymb,amsfonts}
\usepackage{latexsym,verbatim}
\usepackage{bm}
\usepackage{bbold}
\usepackage{color}
\usepackage{ulem}
\usepackage[breaklinks=false,colorlinks,citecolor=blue,linkcolor=blue,urlcolor=blue]{hyperref}
\usepackage[abs]{overpic}
\usepackage{textcomp} 
\usepackage{mathtools}
\usepackage{braket}
\usepackage{comment}
\usepackage{tcolorbox}
\usepackage{soul}    
\usepackage{acronym}    
\usepackage[capitalise]{cleveref}
\usepackage{physics}

\newcommand\FT[1]{\textcolor{black}{#1}}

\begin{document}

\title{Herzberg-Teller coupling in coherent multidimensional spectroscopy: \\ analytical response functions for multilevel systems}

\date{\today}

\author{Filippo Troiani}
\affiliation{Centro S3, CNR-Istituto di Nanoscienze, I-41125 Modena, Italy}

\begin{abstract}
Coherent multidimensional spectroscopy enables detailed investigations of vibronic effects in molecular and solid-state systems. We present explicit analytical expressions for multidimensional nonlinear response functions in the presence of Herzberg-Teller (non-Condon) coupling, within the displaced harmonic oscillator model. The formulation applies to electronic systems with an arbitrary number $N$ of electronic states and to response functions of arbitrary order $M$ in the light-matter interaction. We show that Herzberg-Teller coupling introduces additional oscillatory factors in the time-domain response functions, leading, upon Fourier transformation, to replicas of the Franck-Condon multidimensional spectra shifted by integer multiples of the vibrational frequencies. The present results provide a general analytical framework for the interpretation of non-Condon effects in coherent multidimensional spectroscopies. {\color{black} A Python code that implements the present approach and simulates multidimensional spectra for $N$-level systems is available on GitHub.}
\end{abstract}

\date{\today}

\maketitle

\section{Introduction}
Coherent multidimensional spectroscopy has emerged as a powerful tool for probing ultrafast photophysical processes in molecular and solid-state systems. The nonlinear optical response depends on the time intervals (waiting times) between sequences of phase-locked laser pulses that interact with the system, thus encoding dynamical information into the measured signal\cite{Mukamel95a,Hamm2011}. As a result, multidimensional spectra establish correlations between excitation and detection frequencies, allowing the disentanglement of ultrafast dynamical processes that take place in a wide range of physical, chemical, and biological systems\cite{Schlau11a,Scholes2011,Smallwood18a,Rozzi18a,Maiuri2020,Herkert21a}.

Particular attention has been devoted to the interplay between electronic and vibrational degrees of freedom, that is, to the coupling between electronic and nuclear motions. Such vibronic interactions play a central role in many photoinduced processes and reactions\cite{Luer2009,Rury2017,Thouin2019,DeSio16a,Pandya2018,Womick2011,Falke14a,Rafiq2021,Monahan2017,Schultz2021,Horstmann2020,Scholes2017,Collini2019,Collini2021}. In the adiabatic limit, the light-matter interactions induce purely electronic transitions, with transition dipole moments that are independent of the nuclear coordinates. Nuclear motion nonetheless modulates the optical response through the overlap of vibrational wave packets evolving on different potential energy surfaces, as captured by the vibrational component of the response function\cite{Quintela22a}.

Beyond the adiabatic approximation, non-Condon effects can be incorporated by expanding the electric dipole moment operator $\hat{\mu}$ (projected along the field polarization) in terms of the nuclear normal coordinates $Q$. Retaining the terms linear in $Q$ introduces contributions proportional to the derivatives of the dipole operator with respect to the nuclear coordinates.  These terms give rise to the Herzberg–Teller (HT) coupling, which can originate from nonadiabatic couplings between electronic states associated with distinct potential energy surfaces.

The Herzberg-Teller coupling plays a pivotal role in a broad range of physicochemical phenomena. It influences electron and charge transfer rates\cite{Yoneda20a,Lombardi86a,Kambhampati00a,Birke20a,Soudackov15a,Yu22a,Prakash20a}, radiative and non-radiative decay pathways\cite{Manian21a,Niu10a,Yin20a}, vibronic-polaritonic light emission\cite{Avramenko22a,Fischer23a}, and electronic–vibrational energy transfer\cite{Arsenault21a,Zhang16a,Kong21a}. HT coupling also contributes to internal conversion processes\cite{Lombardi22a,Valiev19a,Valiev20a,Valiev23a,Bracker21a} and shapes non-Condon effects observed in chromophores\cite{Manian22a} and light-harvesting complexes \cite{Arsenault21a}. Interference between the Franck-Condon (FC) and HT contributions can lead to asymmetric and temperature-dependent spectral features in metal-based tetrapyrroles \cite{Roy22a} and modify excited-state Raman and hyper-Raman responses\cite{Chandran24a}. Analytical solutions of two-dimensional response functions based on the use of two electronic levels and on displaced harmonic oscillator Hamiltonians\cite{Allan25a,Allan25b} have shown that FC–HT interference generates asymmetric intensity modulations in absorption spectra\cite{Kundu22a}. These findings have been validated by temperature-dependent studies in systems such as crystalline pentacene, where HT coupling was shown to be essential to reproduce experimental spectra\cite{Qian20a}.

In this work, we present explicit analytical solutions for multidimensional response functions in the presence of HT coupling, within the normal mode approximation. The expressions we derive apply to the displaced harmonic oscillator model of a generic $N$-level system, and to contributions of arbitrary order $M$ in the light-matter interaction. The solution scheme is based on the reduction of the response functions in the presence of HT coupling to those that apply within the FC approximation, whose general expressions we had previously derived through a coherent-state representation of the vibrational states. Within the FC framework, the response function is factored into an electronic and a vibrational contribution, the latter taking the form of a double exponential function of the waiting times\cite{Quintela22a}. 

We show that the inclusion of Herzberg–Teller coupling leads to response functions that can be expressed as the product of the corresponding FC response functions and additional complex oscillatory functions of the waiting times. Upon Fourier transformation with respect to these times, the multidimensional spectra consist of a series of replicas of the full FC spectrum, each shifted in frequency space by integer multiples of the vibrational frequencies. At the spectral level, HT coupling therefore manifests itself through the superposition of multiple replicas of the FC multidimensional spectrum—including its vibrational sidebands—with amplitudes determined by the relevant transition dipole moment matrix elements.

The remainder of the paper is organized as follows. In Sec. \ref{s2} we specify the models, based on the displaced harmonic oscillator Hamiltonian, and outline the solution scheme. In Sec. \ref{s3} we present the main results, namely the expressions of the response functions in a closed (for $M\le 3$) or in a recursive form ($M>3$). In Sec. \ref{s4} we discuss the effect of the HT coupling on the multidimensional spectra of the systems. Finally, the conclusions are drawn in Sec. \ref{s5}. All the derivations are given in the Appendices, which also provide all the intermediate functions used in the previous sections. 

\section{Model and method\label{s2}}

The results derived in the present paper are based on the linearly displaced harmonic oscillator model\cite{Mukamel95a}. This essentially assumes that each vibrational mode of the system can be modeled as an independent harmonic oscillator, whose equilibrium position depends on the electronic state. The model is defined by the following Hamiltonian:
\begin{eqnarray}\label{eq:hamiltonian}
    \hat{H} &=& \sum_{j=0}^{N-1} |j \rangle\langle j| \otimes ( \epsilon_j + \hat{H}_{v,j}) ,
\end{eqnarray}
where $|j\rangle$ and $\epsilon_j$ are the eigenstates and eigenvalues of the electronic Hamiltonian, respectively. The electronic-state dependent vibrational Hamiltonian reads:
\begin{eqnarray}
    \hat{H}_{v,j} &=& \hbar\omega (\hat{a}^\dagger + z_j)(\hat{a} + z_j)\,,
\end{eqnarray}
where $\hat{a}$ and $\hat{a}^\dagger$ are the annihilation and creation operators acting on the vibrational mode state, and the real parameter $z_j$ is the electronic-state dependent displacement of the position operator $\hat{X}=\frac{1}{2}(\hat{a}+\hat{a}^\dagger)$. Without loss of generality, the displacement corresponding to the electronic ground state is set ot zero ($z_0=0$).

Within the perturbative approach to the light-matter interaction, the response function is expanded in different terms, each accounting for a given order $M$. The $M$-th order response function corresponds to the expectation value of an operator, given by nested commutators of the dipole operators\cite{Hamm2011}. Hereafter, we focus on one of the $2^M$ contributions that result from the expansion of the commutators, because all the other ones can be obtained from the first one by a suitable transformation between the waiting times\cite{Troiani23a}. This term, corresponding to a double-sided Feynman diagram where all the light-matter interactions take place on the left side, reads 
\begin{gather}
    \langle \hat{\mu}(t_1+\dots+t_M)\,\hat{\mu}(t_1+\dots+t_{M-1})\dots\,\hat{\mu}(t_1)\,\hat{\mu}(0)\hat{\rho}(0)\rangle \nonumber\\
    =\langle\psi_0| \hat{\mathcal{U}}^\dagger(t_{1M})\,\hat{\mu}\,\hat{\mathcal{U}}(t_M)\,\hat{\mu}\,\hat{\mathcal{U}}(t_{M-1})\dots\hat{\mu}\,\hat{\mathcal{U}}(t_1)\,\hat{\mu}|\psi_0\rangle\!\equiv\! G\,,\label{eq92}
\end{gather}
where $t_{1M}\equiv\sum_{k=1}^M t_k$.
Here, the time-dependent dipole operators $\hat{\mu}$ are given in the Heisenberg picture,
$\hat{\mu}=\hat{\mu}(0)$ corresponds to the dipole operator in the Schr\"odinger picture, $\hat{\rho}(0)=|\psi_0\rangle\langle\psi_0|$ is the initial state, and 
\begin{gather}
    \hat{\mathcal{U}} (t) = e^{-i\hat{H}t} = \sum_j |j \rangle\langle j| \otimes [e^{-i\epsilon_j t}+\hat{U}_j(t)]
\end{gather}
is the time-evolution operator, being $\hat{U}_j(t)\equiv e^{-i\hat{H}_{v,j}t}$ and, from here onward, $\hbar \equiv 1$. In the following, we consider initial states $|\psi_0\rangle = |0,\alpha\rangle$, where the electronic and vibrational degrees of freedom are initialized in the ground state and in a generic coherent state, respectively.

The dipole operator, linear in the annihilation and creation operators, reads:
\begin{gather}\label{eq93}
    \hat{\mu} = \hat{\mu}_0 \otimes \hat{\mathcal{I}} + \hat{\mu}_1 \otimes (\hat{a}+\hat{a}^\dagger) \,,
\end{gather}
where $\hat{\mu}_0$ is the nuclear-coordinate independent component of the dipole matrix element along the direction of the electric field, and $\hat{\mu}_1$ is its first-order derivative with respect to this coordinate {\color{black} (up to a prefactor $1/2$)}. Both operators act on the electronic states, whereas $\hat{\mathcal{I}}$ defines the identity operator in the vibrational space. 

In the FC approximation, the dipole operator is identified with the first term in the above equation. The propagator in Eq. (\ref{eq92}) can thus be decomposed into a number of contributions, each corresponding to a sequence of electronic states $e_0,e_1,\dots,e_M,e_{M+1}$, where $e_0=e_{M+1}=0$ and the transition between $e_{k-1}$ and $e_{k}$ is induced by the $k$-th dipole operator from the left in the expression of $G$. From this it follows that:
\begin{gather}
    G_0 \equiv \langle\psi_0| \hat{\mathcal{U}}^\dagger(t_{1M})\,\hat{\mu}_0\,\hat{\mathcal{U}}(t_M)\,\hat{\mu}_0\dots\hat{\mu}_0\,\hat{\mathcal{U}}(t_1)\,\hat{\mu}_0|\psi_0\rangle\nonumber\\=\!\sum_{\bf e} C_{\bf 0}^{\bf e} \langle \alpha |{\color{black} \hat U_{e_0}(t_{1M})}\hat{U}_{e_{M}}(t_{\color{black}M})\dots \hat{U}_{e_1}(t_1)|\alpha\rangle\times\nonumber\\ e^{-i\sum_{k=1}^M \epsilon_{e_k}t_k}\equiv {\color{black}\sum_{\bf e}} C_{\bf 0}^{\bf e} \,g^{(e)}_{\bf e}(t_1,\dots,t_M)\,g^{(v,\alpha)}_{\bf e}(t_1,\dots,t_M)\,
\end{gather}
where the term in parentheses accounts for the electronic part, $C_{\bf 0}^{\bf e}=\langle e_{M+1}|\hat{\mu}_0|e_M\rangle\langle e_M|\hat{\mu}_0|e_{M-1}\rangle\dots\langle e_1|\hat{\mu}_0|e_0\rangle$, ${\bf e}=(e_1,\dots,e_M)$, ${\bf 0}$ is a vector with $M+1$ zero components, and the factor on the right-hand side corresponds to the vibrational part of the propagator. The expression of $G_0$ [from the following Section referred to as $g_{\bf e}^{(e)}g_{\bf e}^{(v,\alpha)}r_{{\bf e},0}$] has been analytically provided in Ref. \onlinecite{Quintela22a} for a model comprising an arbitrary number $N$ of electronic levels and for an arbitrary order $M$ of the interaction with the electromagnetic field, starting from a coherent-state representation of the vibrational degrees of freedom.  

The HT contributions, which represent the specific target of the present investigation, result from the presence of the second component of the dipole operator [Eq. (\ref{eq93})]. Whenever an operator $\hat{\mu}_0$ that appears in the expression of $G_0$ is replaced by $\hat{\mu}_1(\hat{a}+\hat{a}^\dagger)$, the factor $\langle e_k | \hat{\mu}_0 |e_{k-1}\rangle$ in the expression of $C_{\bf 0}^{\bf e}$ has to be replaced with $\langle e_k | \hat{\mu}_1 |e_{k-1}\rangle$, and the product $\hat{U}_{e_k}(t_{k})\,\hat{U}_{e_{k-1}}(t_{k-1})$ with $\hat{U}_{e_k}(t_{k})\,(\hat{a}+\hat{a}^\dagger)\,\hat{U}_{e_{k-1}}(t_{k-1})$. This implies that, while in the FC approximation the time evolution of the vibrational state
can be regarded as a sequence of rotations of a coherent state around different points in the phase space (each induced by an operator $\hat{U}_{e_k}(t_{k})$)\cite{Quintela22a}, this is no longer the case in the presence of HT contributions, because of the presence of the operator $ (\hat{a}+\hat{a}^\dagger)$ between one time-evolution operator and the other. One of the main findings of the present paper is that, despite this complication, the analytical expression of the vibrational propagator can be written as a combination of overlaps between vibrational coherent states, and its expression is given by the product of the FC propagator $g^{(v,\alpha)}_{\bf e}(t_1,\dots,t_M)$ and of a combination of simple oscillating functions (see Sec. \ref{s3} and Appendices B to D). This result is obtained by progressively moving the creation (annihilation) operators to the left (right) in the expression of the propagator $G$, and finally applying them to the initial {\color{black} coherent} state of the vibrational mode. This, in turn, is achieved by exploiting the commutation relations between the creation and annihilation operators on the one hand and the time-evolution operators $\hat{U}_{e_k}(t_{k})$ on the other (Appendix A).

\section{Results\label{s3}}

This Section contains the main results of the present article, namely the analytical expressions of the $M$-th order response functions for systems with $N$ electronic levels. We show that the overall response functions can always be written as the product of three terms: the electronic response function\cite{Hamm2011}; the vibrational response function corresponding to the FC approximation\cite{Quintela22a}; a third term, which accounts for the HT corrections and which consists of a combination of complex exponential functions of the waiting times. Explicit solutions are provided for the first-, second-, and third-order response functions ($M\le 3$) in terms of intermediate functions, whose dependence on the complex exponential functions is given in the Appendices. For the case of higher-order response functions ($M>3$), analytical solutions are provided in a recursive form. 

{\color{black}The contributions to each response function are divided on the basis of the power $n$ with which the operator $\hat{\mu}_1$ appears. In order to avoid confusion between $n$ and the order in the light-matter interaction ($M$), we refer to the above contributions as {\it linear} ($n\!=\!1$), {\it quadratic} ($n\!=\!2$), {\it cubic} ($n\!=\!3$), and {\it quartic} ($n\!=\!4$) {\it HT terms}. Unless differently specified, the expressions {\it first-}, {\it second-}, {\it third-order} refer instead to the perturbative expansion in the light-matter interaction ($M\!=\!1$, $M\!=\!2$, and $M\!=\!3$, respectively).}

\subsection{First-order response functions}

\begin{figure}
    \centering
    \includegraphics[width=0.95\linewidth]{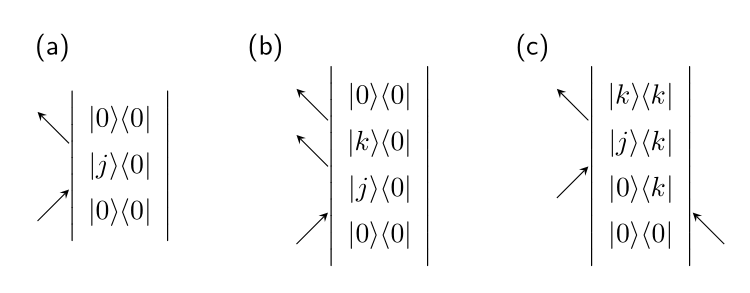}
    \caption{Double-sided Feynman diagrams corresponding to (a) first- and (b,c) second-order processes; $|j\rangle$ and $|k\rangle$ represent the involved electronic states, with $j>k$. }
    \label{fig1}
\end{figure}

The first-order processes correspond to the case where the first interaction with the electric field creates a coherence between the electronic ground state $|0\rangle$ and the excited state $|j\rangle$, followed by a waiting time $t_1$ [Fig. \ref{fig1}(a)].
The general expression of the response function can be written in the following form:
\begin{gather}\label{eq13}
    R^{(1)}_{j,\alpha} (t_1) = ig_j^{(e)} g_j^{(v\FT{,\alpha})} \sum_{p=0}^{2} r_{j,p} \,.
\end{gather}
The function $g_j^{(e)} \equiv e^{-i\epsilon_j t_1}$ (being $\epsilon_j$ the energy of the electronic state $|j\rangle$ {\color{black} and $\hbar\equiv 1$}) defines the electronic component of the response function \cite{Hamm2011}. \FT{If the initial state is a coherent state $|\alpha\rangle$,} the vibrational component is given by\cite{Quintela22a}: 
\begin{gather}\label{eq89}
    g_j^{(v\FT{,\alpha})} = \exp\left[ z_j^2 \left( \chi_1 - 1\right) \FT{+i\Delta\varphi} \right]\,,
\end{gather}
where $\chi_k\equiv e^{-i\omega t_k}$ \FT{and $\Delta\varphi = 2z_j{\rm Im}[\alpha(\chi_1-1)]$}, together with the terms {\color{black}$r_{j,p}$} in the summation of Eq. (\ref{eq13}). 

The first contribution to the summation is the constant
\begin{gather}
    r_{j,0} = C_{00}^j\,,
\end{gather}
corresponding to the FC term, which can be expressed as the product of dipole matrix elements [Eq. (\ref{eq14})]. The following contributions, which account for the HT corrections, are given by terms that oscillate in time at the vibrational frequency $\omega$. In particular, the {\color{black}linear and quadratic HT} contributions read as follows:\begin{gather}
    r_{j,1} = C_{01}^j (f_{|1}^j+f_{1|}^j) + C_{10}^j (f_{2|}^j+f_{|2}^j)   
\end{gather}
and
\begin{gather}
    r_{j,2} = C_{11}^j (f_{12|}^j+f_{1|2}^j+f_{2|1}^j+f_{|12}^j) \,, 
\end{gather}
respectively. The former one accounts for FC-HT interference effects and the latter one for purely HT contributions. \FT{In the notation adopted for the functions $f^{j\dots}_{k\dots|l\dots}$, the superscript specifies the electronic state(s) involved in the propagators, while the numbers on the left- and right-hand sides of the vertical bar in the subscript specify the positions of the annihilation and creation operators, respectively (see Appendix B).} The expression of the coefficients $C_{pq}^j$ is given in Eq. (\ref{eq14}), that of the functions $f$ is given in Eqs. (\ref{eq01p}-\ref{eq09}).
The derivation of the above equations is given in Appendix B.

\subsection{Second-order response functions}

We start by considering second-order processes where the first two interactions with the electric field create two different electronic coherences, between $|0\rangle$ and the electronic excited states $|j\rangle$ and $|k\rangle$. The two interactions are followed by two waiting times, $t_1$ and $t_2$ [Fig. \ref{fig1}(b)]. We note that these second-order contributions can only be nonzero if the system lacks inversion symmetry \cite{Hamm2011}. 

As in the previous Subsection, we write the overall expression of the response function in the form: 
\begin{gather}\label{eq18}
    R^{(2)}_{jk,\alpha} (t_1,t_2) = g_{jk}^{(e)} g_{jk}^{(v\FT{,\alpha})} \sum_{p=0}^{3} r_{jk,p}\,.
\end{gather}
In the case where all interactions with the field occur on the left side of the Feynman diagram [panel (b)], the electronic component of the response function is $ g_{jk}^{(e)} = e^{-i(\epsilon_j t_1+\epsilon_k t_2)}$ (being {\color{black}$\hbar\equiv 1$, and} $\epsilon_j$ and $\epsilon_k$ the energies of the electronic states $|j\rangle$ and $|k\rangle$, respectively). \FT{For $\alpha=0$} the vibrational component is given by\cite{Quintela22a}: 
\begin{gather}\label{eq99}
    g_{jk}^{(v)} \!=\! \exp\left[z_jz_{jk} (\chi_1\!-\!1)\!+\!z_{kj}z_k(\chi_2\!-\!1)\!+\!z_jz_k(\chi_{12}\!-\!1)\right]\,,
\end{gather}
with $\chi_{kl}\equiv e^{-i\omega \sum_{j=k}^l t_j}$, together with the terms in the summation of Eq. (\ref{eq18}). \FT{For $\alpha\neq 0$ the above expression becomes $g_{jk}^{(v,\alpha)} = g_{jk}^{(v)}e^{i\Delta\varphi}$, where $\Delta\varphi = 2\sum_{p=1}^2 z_{e_p}{\rm Im} [\alpha(\chi_p-1)\chi_{1,p-1}]$.}

The first contribution in this summation is the constant
\begin{gather}
    r_{jk,0} = C_{000}^{jk}\,,
\end{gather}
which represents the FC term [Eq. (\ref{eq19})]. 
The {\color{black}linear} HT contribution, which accounts for FC-HT interference effects, is given by the function:{\color{black}
\begin{gather}
    r_{jk,1} = C_{001}^{jk} (f_{1|}^{jk}+f_{|1}^{jk}) + C_{010}^{jk} (f_{2|}^{jk}+f_{|2}^{jk}) \nonumber\\+ C_{100}^{jk} (f_{3|}^{jk}+f_{|3}^{jk}) \,.
\end{gather}}
The above $f$ functions, and thus $r_{jk,1}$, can be expressed in terms of the $\chi_{j,k}$ and $f_{k,l}$ functions. In order to avoid lengthy equations, we do not report here explicitly such expressions, which are provided in Eqs. (\ref{eq27}-\ref{eq30}). 
The {\color{black}quadratic} HT contribution, which also accounts for FC-HT interference effects, is given by the function:{\color{black}
\begin{gather}
    r_{jk,2} = C_{011}^{jk} ( f_{|12}^{jk} + f_{2|1}^{jk}  + f_{1|2}^{jk}  + f_{12|}^{jk} ) \nonumber\\+ C_{110}^{jk} (f_{23|}^{jk}+f_{3|2}^{jk}+f_{2|3}^{jk}+f_{|23}^{jk}) \nonumber\\+C_{101}^{jk} (f_{3|1}^{jk}+f_{1|3}^{jk}+f_{13|}^{jk}+f_{|13}^{jk}) \,.
\end{gather}}
The expressions of the relevant functions $f$ are given in Eqs. (\ref{eq31}-\ref{eq35}). 
Finally, the {\color{black}cubic}, purely HT contribution reads:{\color{black}
\begin{gather}
    r_{jk,3} \!=\! C_{111}^{jk} \!\left(\!f_{123|}^{jk} \!+\! \sum_{\{p_1,p_2\}}\! f_{p_1p_2|q_1}^{jk} \!+\! \sum_{\{p_1\}}\! f_{p_1|q_1q_2}^{jk}\! +\! f_{|123}^{jk}\!\right)\! .
\end{gather}
In the above sums, the indices $(p_1,p_2,q_1)$ and $(p_1,q_1,q_2)$ run over all permutations of $(1,2,3)$ with $p_k<p_{k+1}$ and $q_k<q_{k+1}$.}
The $f$ functions that enter the above equation are given in Eqs. (\ref{eq36},\ref{eq37}), 
the coefficients $C^{jk}_{abc}$ are given in Eq. (\ref{eq19}). 
The derivation of the above equations is provided in Appendix C.

The Feynman diagram reported in Fig. \ref{fig1}(c) is characterized by the same sequence of electronic states and of interactions with the field (arrows), read in clockwise order from the bottom left to the bottom right of the diagram, as that reported in Fig. \ref{fig1}(b). The corresponding response function can thus be obtained from the above $R^{(2)}_{jk,\alpha}$ by a suitable transformation between the waiting times, following a procedure that is discussed in detail in Ref. \onlinecite{Troiani23a}. In this case, the transformation is given by $t_1 \rightarrow t_2$ and $t_2 \rightarrow -t_1-t_2$. This implies that: 
\begin{gather}\label{eq39p}
\chi_1 \rightarrow \chi_2\,,\ \ \chi_2 \rightarrow \chi_{12}^*\,,
\end{gather}
and therefore $\chi_{12} = \chi_1 \chi_2\rightarrow \chi_{1}^*$.
As a result, the electronic component reads $ g_{jk}^{(e)} = e^{-i[\epsilon_j t_2-\epsilon_k (t_1+t_2)]}$, where {\color{black} $\hbar\equiv 1$. In addition, one should add a minus sign} in front of the exponential function, {\color{black}resulting} from the modified number of arrows on the right side of the Feynman diagram\cite{Hamm2011}. As to the vibrational component of the response function, \FT{for $\alpha=0$} its expression becomes:
\begin{gather}\label{eq39}
    g_{jk}^{(v)} \!=\! \exp\left[z_jz_{jk} (\chi_2\!-\!1)\!+\!z_{kj}z_k(\chi^*_{12}\!-\!1)\!+\!z_jz_k(\chi_{1}^*\!-\!1)\right]\,.
\end{gather}
The expressions of the HT contributions (functions $r_{jk,p}$, with $p=1,2,3$) can be obtained from the ones above through the replacements specified in Eq. (\ref{eq39p}), to be applied also within the functions $f_{k,l}$. Finally, the coefficients $C_{abc}^{jk}$ remain unchanged.

\subsection{Third-order response functions\label{subsec:torf}}

\begin{figure}
    \centering
    \includegraphics[width=0.95\linewidth]{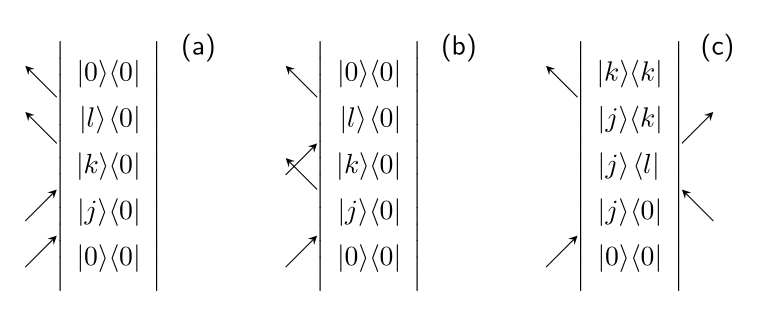}
    \caption{Double-sided Feynman diagrams corresponding to third-order processes; $|j\rangle$, $|k\rangle$, and $|l\rangle$ represent the involved electronic states, with $k>j,l$ in (a), $k<j,l$ in (b,c). }
    \label{fig2}
\end{figure}

We start by considering third-order processes where the first three interactions with the electric field create three different electronic coherences, each one followed by one waiting time, $t_1$, $t_2$, and $t_3$ [Fig. \ref{fig2}(a)]. The general expression of the third-order response function reads as follows:
\begin{gather}\label{eq87}
    R^{(3)}_{jkl,\alpha} (t_1,t_2,t_3) = -ig_{jkl}^{(e)} g_{jkl}^{(v\FT{,\alpha})} \sum_{p=0}^{4} r_{jkl,p} \,.
\end{gather}
The function $ g_{jkl}^{(e)} \equiv e^{-i(\epsilon_j t_1+\epsilon_k t_2+\epsilon_l t_3)}$ is the electronic component of the response function (being {\color{black}$\hbar\equiv 1$,} $\epsilon_p$ the energy of the electronic state $|p\rangle$, with $p=j,k,l$). {\color{black}For $\alpha=0$} the vibrational component is given by
\begin{gather}
    g_{jkl}^{(v)} = \exp\left[z_{j}z_{jk}(\chi_1-1)\right.\nonumber\\ +z_{kj}z_{kl}(\chi_2-1)+z_{lk}z_{l}(\chi_3-1)+z_{j}z_{kl}(\chi_{12}-1)\nonumber\\ \left. +z_{kj}z_{l}\,(\chi_{23}-1)+z_{j}z_{l}(\chi_{13}-1) \right] \equiv e^f\label{eq91}\,,
\end{gather}
together with the terms in the summation of Eq. (\ref{eq87}). \FT{For $\alpha\neq 0$ the above expression becomes $g_{jk}^{(v,\alpha)} = g_{jk}^{(v)}e^{i\Delta\varphi}$, where $\Delta\varphi = 2\sum_{p=1}^3 z_{e_p}{\rm Im} [\alpha(\chi_p-1)\chi_{1,p-1}]$.}

The first contribution in this summation is the constant 
\begin{gather}\label{eqy01}
    r_{jkl,0} = C_{0000}^{jkl}\,,
\end{gather}
which represents the FC term [Eq. (\ref{eq38})]. The {\color{black}linear} HT contribution, which accounts for FC-HT interference effects, is given by the function:{\color{black}
\begin{gather}
    r_{jkl,1} = C_{0001}^{jkl} (f_{1|}^{jkl}+f_{|1}^{jkl}) + C_{0010}^{jkl} (f_{2|}^{jkl} + f_{|2}^{jkl}) \nonumber\\ + C_{0100}^{jkl} (f_{3|}^{jkl} + f_{|3}^{jkl} ) + C_{1000}^{jkl} (f_{4|}^{jkl}+f_{|4}^{jkl}) \,.\label{eqx01}
\end{gather}}
The $f$ functions that appear in the above equation can be expressed in terms of the $\chi_{j,k}$ and $f_{k,l}$ functions [Eqs. (\ref{eq54}-\ref{eq59})]. 
The {\color{black}quadratic} HT contribution is given by the function:{\color{black}
\begin{gather}
    r_{jkl,2} = C_{0011}^{jkl} \left(f_{12|}^{jkl} + f_{1|2}^{jkl} + f_{2|1}^{jkl} + f_{|12}^{jkl} \right) + \nonumber\\ C_{0101}^{jkl} \left(f_{13|}^{jkl} + f_{1|3}^{jkl} + f_{3|1}^{jkl} + f_{|13}^{jkl}\right) \nonumber\\ + C_{1001}^{jkl} \left(f_{14|}^{jkl} + f_{1|4}^{jkl} + f_{4|1}^{jkl} + f_{|14}^{jkl}\right) \nonumber\\ + C_{0110}^{jkl} \left(f_{23|}^{jkl} + f_{2|3}^{jkl} + f_{3|2}^{jkl} + f_{|23}^{jkl} \right) \nonumber\\ + C_{1010}^{jkl} \left(f_{24|}^{jkl} + f_{2|4}^{jkl} + f_{4|2}^{jkl} + f_{|24}^{jkl} \right) \nonumber\\ + C_{1100}^{jkl} \left( f_{34|}^{jkl} + f_{3|4}^{jkl} + f_{4|3}^{jkl} + f_{|34}^{jkl} \right)\,.\label{eqz01}
\end{gather}}
The expressions of the functions $f$ are reported in Eqs. (\ref{eq60}-\ref{eq72}). The {\color{black}cubic} HT contribution, which, like the previous one, accounts for FC-HT interference effects, is given by the equation:{\color{black}
\begin{gather}
    r_{jkl,3} \!=\! 
     C_{1110}^{jkl} \left(f_{234|}^{jkl} \!+\! \sum_{\{p_1,p_2\}} f_{p_1p_2|q_1}^{jkl} \!+\! \sum_{\{p_1\}} f_{p_1|q_1q_2}^{jkl} \!+\! f_{|234}^{jkl}\right)\nonumber\\
     C_{1101}^{jkl} \left(f_{134|}^{jkl} \!+\! \sum_{\{p_1,p_2\}}' f_{p_1p_2|q_1}^{jkl} \!+\! \sum_{\{p_1\}}' f_{p_1|q_1q_2}^{jkl} \!+\! f_{|134}^{jkl}\right)\nonumber\\
     C_{1011}^{jkl} \left(f_{124|}^{jkl} \!+\! \sum_{\{p_1,p_2\}}'' f_{p_1p_2|q_1}^{jkl} \!+\! \sum_{\{p_1\}}'' f_{p_1|q_1q_2}^{jkl} \!+\! f_{|124}^{jkl}\right)\nonumber\\
     C_{0111}^{jkl} \left(f_{123|}^{jkl} \!+\! \sum_{\{p_1,p_2\}}''' f_{p_1p_2|q_1}^{jkl} \!+\! \sum_{\{p_1\}}''' f_{p_1|q_1q_2}^{jkl} \!+\! f_{|123}^{jkl}\right)
     \,. 
\end{gather}
In the unprimed sums, the indices $(p_1,p_2,q_1)$ and $(p_1,q_1,q_2)$ run over all permutations of $(2,3,4)$ with $p_k<p_{k+1}$ and $q_k<q_{k+1}$. The same applies for the primed, double and triple primed sums, where $(p_1,p_2,q_1)$ and $(p_1,q_1,q_2)$ are permutations of $(1,3,4)$, $(1,2,4)$, and $(1,2,3)$, respectively.}
The explicit expression of the above $f$ functions is reported in Eqs. (\ref{eq73}-\ref{eq82}).
Finally, the {\color{black}quartic}, purely HT contributions read:{\color{black}
\begin{gather}
    r_{jkl,4} = C_{1111}^{jkl} 
    \left( f_{1234|}^{jkl}+
    \sum_{\{p_1,p_2,p_3\}} f_{p_1p_2p_3|q_1}^{jkl}+\right.\nonumber\\ \left.
    \sum_{\{p_1,p_2\}} f_{p_1p_2|q_1q_2}^{jkl}+
    \sum_{\{p_1\}} f_{p_1|q_1q_2q_3}^{jkl}   +f_{|1234}^{jkl} \right)\,,
\end{gather}
which is completed by Eqs. (\ref{eq83}-\ref{eq86}). In the above sums, the indices $(p_1,p_2,p_3,q_1)$, $(p_1,p_2,q_1,q_2)$, and $(p_1,q_1,q_2,q_3)$ run over all permutations of $(1,2,3,4)$ with $p_k<p_{k+1}$ and $q_k<q_{k+1}$.} The coefficients $C^{jkl}_{abcd}$ are given in Eq. (\ref{eq38}). The derivation of the above equations is given in Appendix D.

\subsection{Higher-order response functions}

The procedure that has been applied above to the derivation of $M$-th order response functions with $M\le 3$ is general and can also be applied to higher-order response functions. It consists of the following steps: 

$(1)$ Calculation of the function $g_{jk\dots}^{(e)}g_{jk\dots}^{(v,\alpha)}$, which corresponds, up to a multiplicative constant, to the result obtained within the FC approximation. The calculation of the electronic part $g_{jk\dots}^{(e)}$ is straightforward\cite{Hamm2011}. As to the vibrational component $g_{jk\dots}^{(v,\alpha)}$, one can use the general solution provided in Ref. \onlinecite{Quintela22a} [Eq. (72)]. In fact, the response functions corresponding to $\alpha=0$ and to a Feynman diagram where all interactions occur on the left side can be expressed as $g_{jk\dots}^{(v)}\equiv e^{f}$, where
\begin{gather}
f \!=\! \sum_{k=0}^{M-1} \sum_{j=1}^{M-k} (z_{e_j}\!\!-\!z_{e_{j-1}})(z_{e_{j+k}}\!\!-\!z_{e_{j+k+1}}) (\chi_{j,j+k}\!-\!1).
\end{gather}
Here, $|e_j\rangle$ is the electronic state during the $j$-th waiting time, $e_0=e_{M+1}=0$, and $z_{e_0}=z_{e_{M+1}}=0$. One can easily verify that Eqs. (\ref{eq89},\ref{eq99},\ref{eq91}) coincide with this expression for $M=1,2,3$. The remaining cases, corresponding to diagrams in which the interactions occur on both sides, can be derived by suitable transformations between the waiting times, as discussed above for $M=2$ and $M=3$. \FT{The generalization of $g^{(v)}$ to the case where the vibrational mode is initialized in a generic coherent state $|\alpha\rangle$, rather than in the ground state (i.e. in the particular coherent state defined by $\alpha = 0$), is given by the equation: $g^{(v,\alpha)}=g^{(v)}e^{i\Delta\varphi}$, where $\Delta\varphi = 2\sum_{p=1}^M z_{e_p}{\rm Im} [\alpha(\chi_p-1)\chi_{1,p-1}]$\cite{Quintela22a}.}

$(2)$ The second step, the first in the derivation of the HT terms, is the calculation of the functions 
\begin{gather}
a^{(M)}_k \equiv 
\frac{\langle \FT{\alpha} | \FT{\hat{U}_{M+1}}\hat{U}_M\dots \hat{U}_k\,\hat{a}\,\hat{U}_{k-1}\dots \hat{U}_1 |\FT{\alpha}\rangle}{\color{black}\langle \alpha | \hat{U}_{M+1}\hat{U}_M\dots \hat{U}_k\,\hat{U}_{k-1}\dots \hat{U}_1 |\alpha\rangle}
\end{gather} 
and 
\begin{gather}
c^{(M)}_k\equiv
\frac{\langle \FT{\alpha} | \FT{\hat{U}_{M+1}}\hat{U}_M\dots \hat{U}_k\,\hat{a}^\dagger\,\hat{U}_{k-1}\dots \hat{U}_1 |\FT{\alpha}\rangle}{\color{black}\langle \alpha | \hat{U}_{M+1}\hat{U}_M\dots \hat{U}_k\,\hat{U}_{k-1}\dots \hat{U}_1 |\alpha\rangle}
\,, 
\end{gather}
\FT{where $\hat{U}_k\equiv e^{-i \hat{H}_{v,e_k}t_k}$}.
These can be expressed in terms of the simple functions $f_{k,l}$ and $\chi_{k,l}$. In fact, from the commutation relations between annihilation or creation and time-evolution operators [Eqs. (\ref{eq04}-\ref{eq02})] it follows that:
\begin{gather}
    a^{(M)}_{k} = a^{(M)}_{k-1} \,\chi_{k-1} + f_{e_{k-1},k-1} \\
    c^{(M)}_k = c^{(M)}_{k+1} \,\chi_k - f_{e_k,k} \,,\label{eq10}
\end{gather}
where $a^{(M)}_{1} = \FT{\alpha}$ and $ c^{(M)}_{M+1} =\FT{\chi^*_{1{\color{black},}M}\,\alpha^*}$. This in turn implies that
\begin{gather}\label{eq88}
    a_k^{(M)} = \sum_{j=1}^{k-1} f_{e_j,j} \,\chi_{j+1,k-1} \,,\ \ 
    c_k^{(M)} = - \sum_{j=k}^M f_{e_j,j} \,\chi_{k,j-1}\,,
\end{gather}
where $\chi_{k,l}\equiv 1$ for {\color{black}$k=l+1$}.

$(3)$ The third step, the second in the derivation of the HT terms, is the calculation of the functions $f_{p_1\dots p_K|q_1\dots q_L}$, corresponding to the expectation values of the operators that consist of the product of the $M$ operators $\hat{U}_k$, $K$ annihilation operators, in positions specified by the indices $p_j$, and the $L$ creation operators, in positions specified by the indices $q_k$. For example, for $M=5$, we have that:
\begin{gather}
    f_{3|24} = \frac{\langle \FT{\alpha} | \FT{\hat{U}_{6}}\hat{U}_5 \hat{U}_4 \,\hat{a}^\dagger\, \hat{U}_3 \,\hat{a}\, \hat{U}_2 \,\hat{a}^\dagger\, \hat{U}_1 |\FT{\alpha}\rangle}{\FT{\langle \alpha | \hat{U}_{6}\hat{U}_5 \hat{U}_4 \,\hat{U}_3 \, \hat{U}_2 \,\hat{U}_1 |\alpha\rangle}} .
\end{gather}

These functions can be reduced to combinations of $a^{(M)}_k$ and $c^{(M)}_k$. The procedure consists in progressively lowering the number of creation operators, through the equation:
\begin{gather}
f_{p_1\dots p_K|q_1\dots q_L} = c_{q_1} f_{p_1\dots p_K|q_2\dots q_L} + \chi_{q_1,p_1-1} f_{p_2\dots p_K|q_2\dots q_L} \nonumber\\ +\! \chi_{q_1,p_2-1} f_{p_1p_3\dots p_K|q_2\dots q_L} \!+\! \dots \!+\! \chi_{q_1,p_K-1} f_{p_1\dots p_{K-1}|q_2\dots q_L},\label{eq90}
\end{gather}
where $\chi_{k,l} = 0$ for {\color{black}$k>l+1$}. When all creations operators have been removed, the remaining function is given by:
\begin{gather}
f_{p_1\dots p_L|} = \prod_{i=1}^K a^{(M)}_{p_i}\,.
\end{gather}
In fact, from Eq. (\ref{eq90}) it follows that the functions corresponding to normally ordered annihilation and creation operators ($p_j<q_k$ for arbitrary $j$ and $k$) can be directly factorized into the product of $a^{(M)}_k$ and $c^{(M)}_k$ functions:
\begin{gather}
f_{p_1\dots p_K|q_1\dots q_L} = \left[\prod_{i=1}^K a^{(M)}_{p_i}\right] \left[ \prod_{j=1}^L c^{(M)}_{q_j}\right]\,.
\end{gather}
As a representative example, let us consider the case where $K=L=2$. If the operators are normally ordered, as for $p_1=1\,, p_2=2\,,q_1=3\,, q_2=4$, then $f_{12|34}=a_1 a_2 c_3 c_4$ (we omit here the superscript $(M)$ for simplicity). If the operators are anti-normally ordered, one starts by reducing the number of creation operators through:
$f_{34|12}=c_1 f_{34|2} + \chi_{12} f_{4|2} + \chi_{13} f_{3|2} $. Applying the relation in Eq. (\ref{eq90}) recursively, one finally obtains: $f_{34|12}=c_1 (a_3 a_4 c_2+ a_3 \chi_{23} + a_4 \chi_2) + \chi_{12} (a_4 c_2 + \chi_{23} ) + \chi_{13} (a_3 c_2 + \chi_2) $.

$(4)$ The last step consists in grouping together the $f_{p_1\dots p_K|q_1\dots q_L}$ functions corresponding to the same HT order ($p=K+L$), and in multiplying them by the respective prefactors $C$. By so doing, one finally obtains the functions $r_{jk\dots,p}(t_1,\dots,t_M)$, which, multiplied by $g_{jk\dots}^{(e)}\,g_{jk\dots}^{(v,\alpha)}$, give the overall response function, as shown in detail for the cases $1\le M \le 3$. 

{\color{black}
\subsection{Further generalizations}

\begin{figure}
    \centering
    \includegraphics[width=0.75\linewidth]{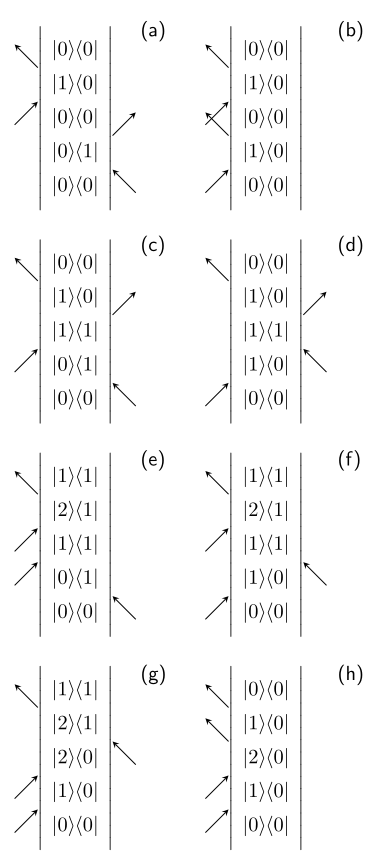}
    \caption{{\color{black}Double-sided Feynman diagrams corresponding to third-order processes ($M\!=\!3$) in a system with three electronic levels: (a,b) ground state bleaching; (c,d) stimulated emission; (e,f) excited state absorption, (g,h) double quantum coherence. }}
    \label{fig3}
\end{figure}

The results presented so far are based on a number of assumptions. These include the presence in the system of a single and harmonic vibrational mode, whose frequency is independent of the electronic state. Also, the Taylor expansion of the electric dipole operator is truncated at the linear term in the nuclear displacement. Finally, the present model does not account for non-adiabatic couplings, which might result in transitions between electronic states mediated by the vibrational motion. As discussed in the remainder of this Section, the present approach can be extended in order to account for finite-temperature effects, for the presence of higher-order terms in the Taylor expansion of the dipole operator, and for contributions related to multiple independent vibrational modes. Extensions of the original method\cite{Quintela22a} have also been carried out that allow the inclusion of Duschinsky mode mixing\cite{Quintela23a} and nonadiabatic effects\cite{Troiani23a}. Nontrivial methodological developments will be needed to combine those developments with the present one and simultaneously account for all these features in a unified manner.

\subsubsection{Finite temperature}

The above equations assume that the vibrational mode is initialized in a coherent state $|\alpha\rangle$. The coherent state $\alpha=0$ coincides with the Fock state $n=0$, and thus gives the response functions in the zero-temperature limit. On the other hand, the coherent states represent an overcomplete basis for the vibrational mode. The finite-temperature response functions can thus be derived by integrating in the phase space, and specifically through: 
\begin{gather}\label{eqwww}
   R^{(M)}_{jk\dots,T}(t_1,\dots,t_M) =  \int\,d^2\alpha\,R^{(M)}_{jk\dots,\alpha}(t_1,\dots,t_M) \,P(\alpha,\alpha^*)\,,
\end{gather}
where $P(\alpha,\alpha^*) = \frac{1}{\pi\langle n \rangle} e^{-|\alpha|^2/\langle n \rangle}$, $1/\langle n\rangle = e^{\hbar\omega/k_BT}-1$, and $\langle n \rangle$ is the average mode occupation. The above equation assumes that the electronic excitation energy is much larger than $k_BT$, so that the temperature only affects the initial state of the vibrational modes. Analytical solutions have been derived in the FC case\cite{Quintela22a}, and are generalized hereafter to the response functions in the presence of HT coupling. 

In the response function $R_{jkl\dots,\alpha}^{(M)}$ [see, e.g., Eq. (\ref{eq87}) for the case $M=3$] the terms that depend on $\alpha$ are the FC response function $g_{jkl\dots}^{(v,\alpha)}=e^{f}$ and most of the contributions to the HT corrections $r_{jkl\dots,p}$. As a result, one has in the expression of $R_{jkl\dots,\alpha}^{(M)}$ a series of terms $\eta_{ab}\equiv\alpha^a(\alpha^*)^b g_{jkl\dots}^{(v,\alpha)} {\color{black}/ g_{jkl\dots}^{(v,0)}} {\color{black}= \alpha^a(\alpha^*)^b e^{i\Delta\varphi}}$, with $a$ and $b$ nonnegative integers. The finite temperature response function results from the above integration [Eq. (\ref{eqwww})], which results in the replacement 
\begin{gather}
\eta_{ab}\!\rightarrow\! I_{ab}\equiv\! \frac{1}{\pi\langle n\rangle}\int d^2\alpha\,\alpha^a(\alpha^*)^be^{-|\alpha|^2/\langle n\rangle} e^{\alpha Q^*-\alpha^*Q}
\nonumber\\= e^{-\langle n\rangle|Q|^2}\! \sum_{k=0}^{\min(a,b)}\! \binom{a}{k}\!\binom{b}{k} k! \, \langle n\rangle^{a+b} (Q^*)^{a-k} Q^{b-k}
\end{gather}
where {\color{black}$i\Delta\varphi=\alpha Q^*-\alpha^* Q$ and} the function $Q$ reads 
\begin{gather}
Q\equiv 2\sum_{j=1}^M z_{e_j}(\chi_j^*-1)\chi^*_{1,j-1}    
\end{gather}
and $e_1=j$, $e_2=k$, {\color{black}$e_3=l$}, etc. 

The expression of $I_{ab}$ takes a simple form for contributions that are of low order in $\alpha$ and $\alpha^*$. In particular, for if one of the two exponents is zero ($ab=0$) and the other is lower than 3 ($a+b < 3$), then
\begin{gather}
    I_{ab}
    = \langle n\rangle^{2(a+b)} (Q^*)^a Q^b e^{-\langle n\rangle |Q|^2} \,.
\end{gather}
This includes the FC result, $I_{00}=e^{-\langle n\rangle |Q|^2}$, as a special case\cite{Quintela22a}. For $a\!=\!b\!=\!1$ the general expression of $I_{ab}$ becomes
\begin{gather}
    I_{ab}\equiv \langle n\rangle^2 [1+\langle n\rangle |Q|^2] e^{-\langle n\rangle |Q|^2} \,.
\end{gather}


\subsubsection{Further expansion of the dipole operator}

\begin{figure*}
    \centering
    \includegraphics[width=0.95\linewidth]{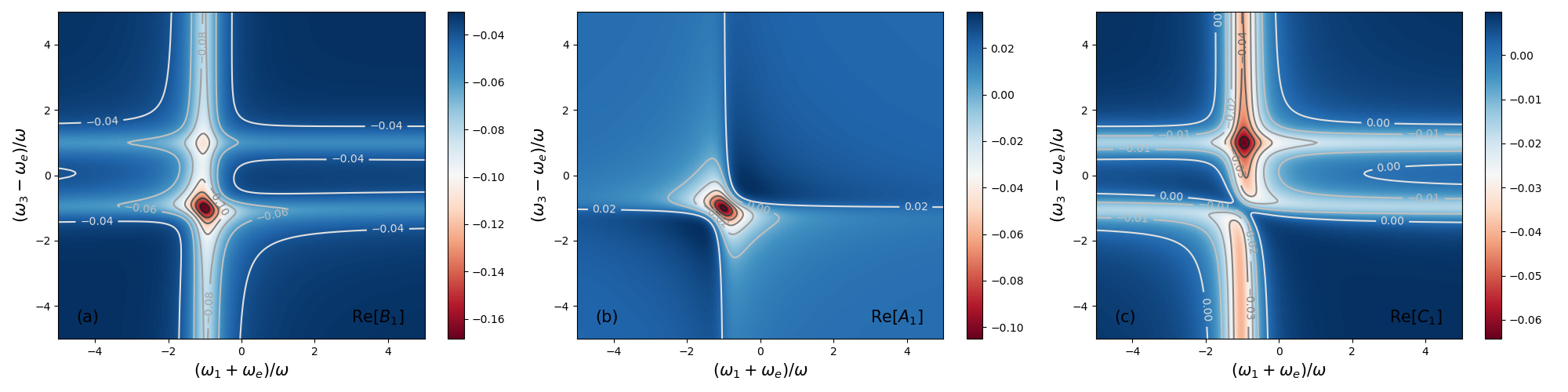}
    \caption{{\color{black}Spectral features associated to the rephasing response function [Fig. \ref{fig3}(a)], ground state bleaching processes, Fourier transformed with respect to $t_1$ and $t_3$, for $t_2=0$. The contour plots refer to the terms in the response function that are: quadratic in $z_1$ and of zero-th order in $\hat{\mu}_1$ [$B_1$, Fourier transform of $g^{(e)}_{101}\tilde{B}_1$, see Eq. (\ref{eqB1}), panel (a)]; linear in $z_1$ and $\hat{\mu}_1$ [$A_1$, Fourier transform of $g^{(e)}_{101}\tilde{A}_1$, see Eq. (\ref{eqA1}), panel (b)]; of zero-th order in $z_1$ and quadratic in $\hat{\mu}_1$ [$C_1$, Fourier transform of $g^{(e)}_{101}\tilde{C}_1$, see Eq. (\ref{eqC1}), panel (c)]. The line broadening is determined by $\gamma=0.4\,\omega$, $\hbar\omega_e=\epsilon_1-\epsilon_0$ is the energy difference between the ground and first excited electronic states.}}
    \label{fGSBr}
\end{figure*}

\begin{figure*}
    \centering
    \includegraphics[width=0.95\linewidth]{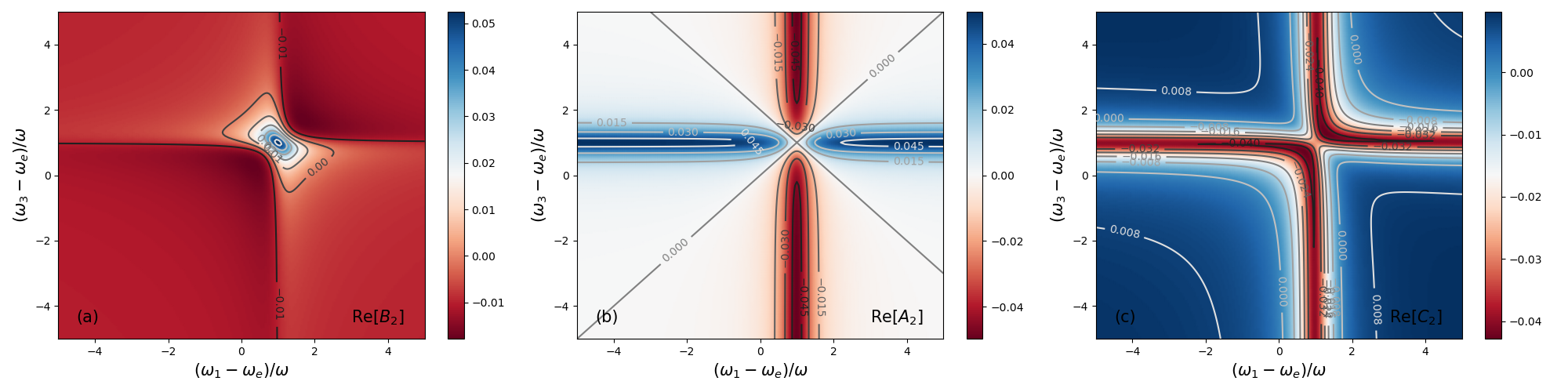}
    \caption{{\color{black}Spectral features associated to the non-rephasing response function [Fig. \ref{fig3}(b)], ground state bleaching processes, Fourier transformed with respect to $t_1$ and $t_3$, for $t_2=0$. The contour plots refer to the terms in the response function that are: quadratic in $z_1$ and of zero-th order in $\hat{\mu}_1$ [$B_2$, Fourier transform of $g^{(e)}_{101}\tilde{B}_2$, see Eq. (\ref{eqB2}), panel (a)]; linear in $z_1$ and $\hat{\mu}_1$ [$A_2$, Fourier transform of $g^{(e)}_{101}\tilde{A}_2$, see Eq. (\ref{eqA2}), panel (b)]; of zero-th order in $z_1$ and quadratic in $\hat{\mu}_1$ [$C_2$, Fourier transform of $g^{(e)}_{101}\tilde{C}_2$, see Eq. (\ref{eqC2}), panel (c)]. The line broadening is determined by $\gamma=0.4\,\omega$, $\hbar\omega_e=\epsilon_1-\epsilon_0$ is the energy difference between the ground and first excited electronic states.}}
    \label{fGSBnr}
\end{figure*}

\begin{figure*}
    \centering
    \includegraphics[width=0.95\linewidth]{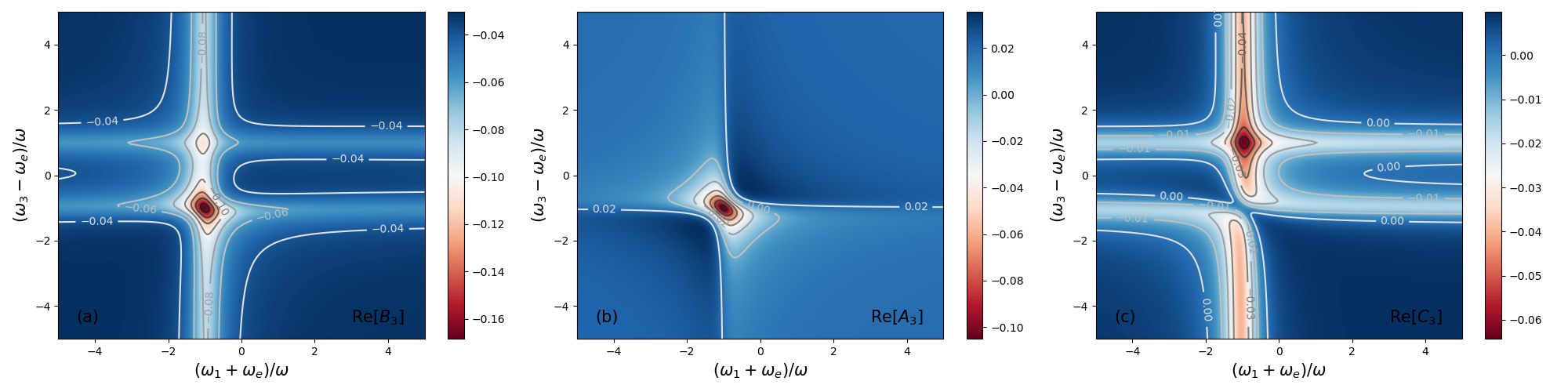}
    \caption{{\color{black}Spectral features associated to the rephasing response function [Fig. \ref{fig3}(c)], stimulated emission processes, Fourier transformed with respect to $t_1$ and $t_3$, for $t_2=0$. The contour plots refer to the terms in the response function that are: quadratic in $z_1$ and of zero-th order in $\hat{\mu}_1$ [$B_3$, Fourier transform of $g^{(e)}_{101}\tilde{B}_3$, see Eq. (\ref{eqB3}), panel (a)]; linear in $z_1$ and $\hat{\mu}_1$ [$A_3$, Fourier transform of $g^{(e)}_{101}\tilde{A}_3$, see Eq. (\ref{eqA3}), panel (b)]; of zero-th order in $z_1$ and quadratic in $\hat{\mu}_1$ [$C_3$, Fourier transform of $g^{(e)}_{101}\tilde{C}_3$, see Eq. (\ref{eqC3}), panel (c)]. The line broadening is determined by $\gamma=0.4\,\omega$, $\hbar\omega_e=\epsilon_1-\epsilon_0$ is the energy difference between the ground and first excited electronic states.}}
    \label{fSEr}
\end{figure*}

\begin{figure*}
    \centering
    \includegraphics[width=0.95\linewidth]{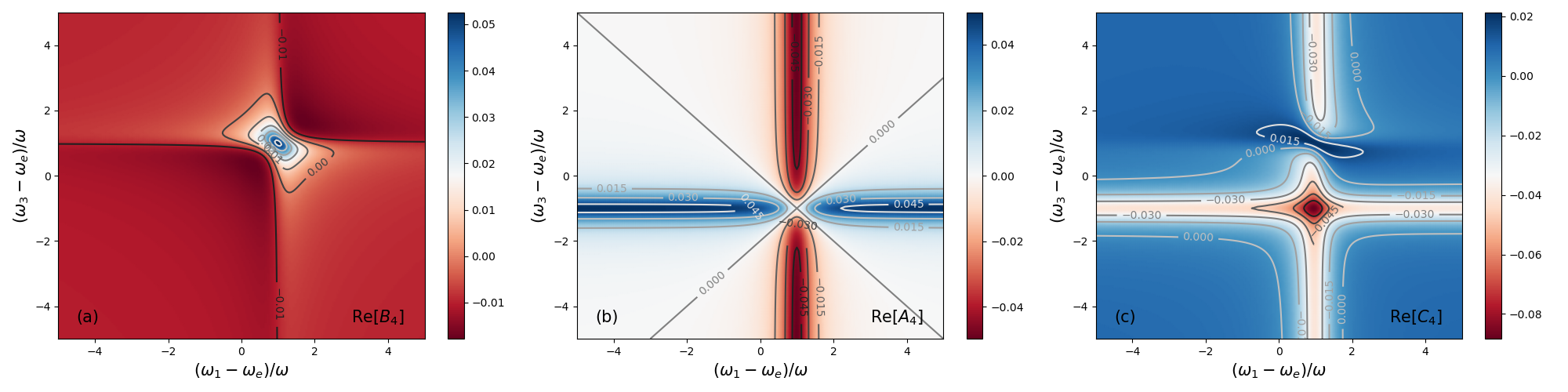}
    \caption{{\color{black}Spectral features associated to the non-rephasing response function [Fig. \ref{fig3}(d)], ground state bleaching processes, Fourier transformed with respect to $t_1$ and $t_3$, for $t_2=0$. The contour plots refer to the terms in the response function that are: quadratic in $z_1$ and of zero-th order in $\hat{\mu}_1$ [$B_4$, Fourier transform of $g^{(e)}_{101}\tilde{B}_4$, see Eq. (\ref{eqB4}), panel (a)]; linear in $z_1$ and $\hat{\mu}_1$ [$A_4$, Fourier transform of $g^{(e)}_{101}\tilde{A}_4$, see Eq. (\ref{eqA4}), panel (b)]; of zero-th order in $z_1$ and quadratic in $\hat{\mu}_1$ [$C_4$, Fourier transform of $g^{(e)}_{101}\tilde{C}_4$, see Eq. (\ref{eqC4}), panel (c)]. The line broadening is determined by $\gamma=0.4\,\omega$, $\hbar\omega_e=\epsilon_1-\epsilon_0$ is the energy difference between the ground and first excited electronic states.}}
    \label{fSEnr}
\end{figure*}

\begin{figure*}
    \centering
    \includegraphics[width=0.95\linewidth]{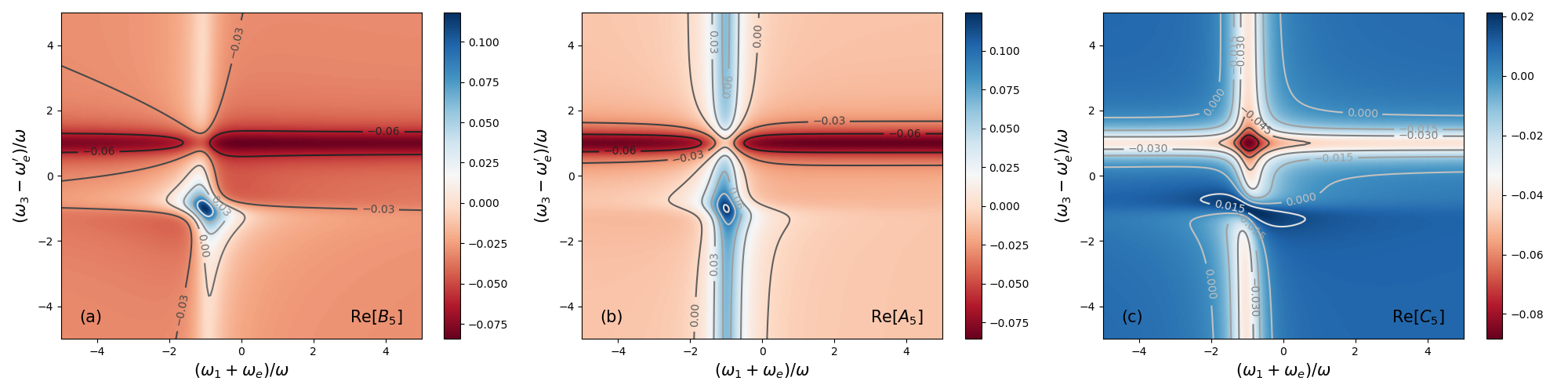}
    \caption{{\color{black}Spectral features associated to the rephasing response function [Fig. \ref{fig3}(e)], excited state absorption processes, Fourier transformed with respect to $t_1$ and $t_3$, for $t_2=0$. The contour plots refer to the terms in the response function that are: quadratic in $z_1$ and of zero-th order in $\hat{\mu}_1$ [$B_5$, Fourier transform of $g^{(e)}_{121}\tilde{B}_5$, see Eq. (\ref{eqB5}), panel (a)]; linear in $z_1$ and $\hat{\mu}_1$ [$A_5$, Fourier transform of $g^{(e)}_{121}\tilde{A}_5$, see Eq. (\ref{eqA5}), panel (b)]; of zero-th order in $z_1$ and quadratic in $\hat{\mu}_1$ [$C_5$, Fourier transform of $g^{(e)}_{121}\tilde{C}_5$, see Eq. (\ref{eqC5}), panel (c)]. The line broadening is determined by $\gamma=0.4\,\omega$, $\hbar\omega_e=\epsilon_1-\epsilon_0$ is the energy difference between the ground and first excited electronic states, whereas $\hbar\omega_e'=\epsilon_2-\epsilon_1$.}}
    \label{fESAr}
\end{figure*}

\begin{figure*}
    \centering
    \includegraphics[width=0.95\linewidth]{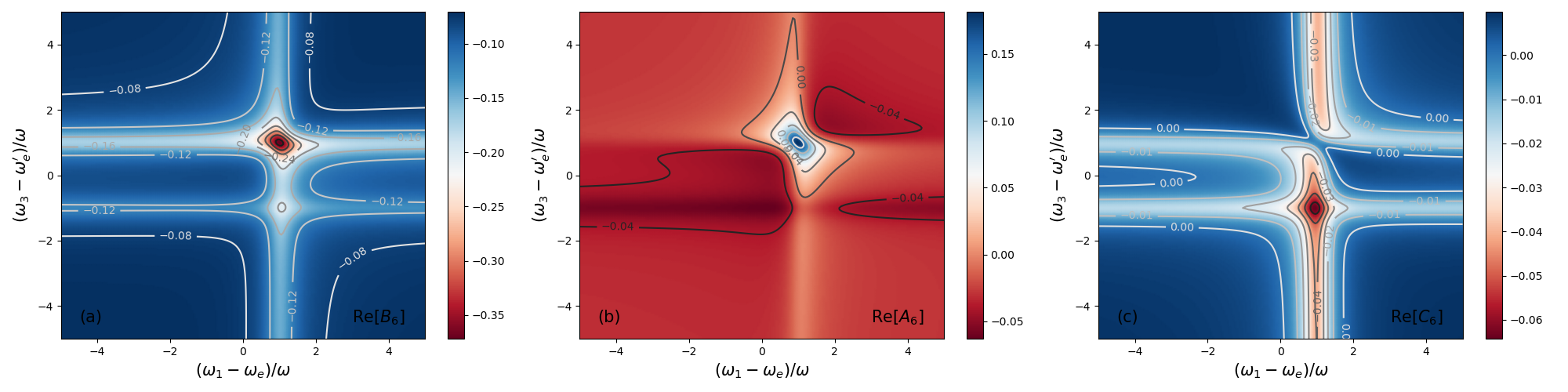}
    \caption{{\color{black}Spectral features associated to the non-rephasing response function [Fig. \ref{fig3}(f)], excited state absorption processes, Fourier transformed with respect to $t_1$ and $t_3$, for $t_2=0$. The contour plots refer to the terms in the response function that are: quadratic in $z_1$ and of zero-th order in $\hat{\mu}_1$ [$B_6$, Fourier transform of $g^{(e)}_{121}\tilde{B}_6$, see Eq. (\ref{eqB6}), panel (a)]; linear in $z_1$ and $\hat{\mu}_1$ [$A_6$, Fourier transform of $g^{(e)}_{121}\tilde{A}_6$, see Eq. (\ref{eqA6}), panel (b)]; of zero-th order in $z_1$ and quadratic in $\hat{\mu}_1$ [$C_6$, Fourier transform of $g^{(e)}_{121}\tilde{C}_6$, see Eq. (\ref{eqC6}), panel (c)]. The line broadening is determined by $\gamma=0.4\,\omega$, $\hbar\omega_e=\epsilon_1-\epsilon_0$ is the energy difference between the ground and first excited electronic states, whereas $\hbar\omega_e'=\epsilon_2-\epsilon_1$.}}
    \label{fESAnr}
\end{figure*}

\begin{figure*}
    \centering
    \includegraphics[width=0.95\linewidth]{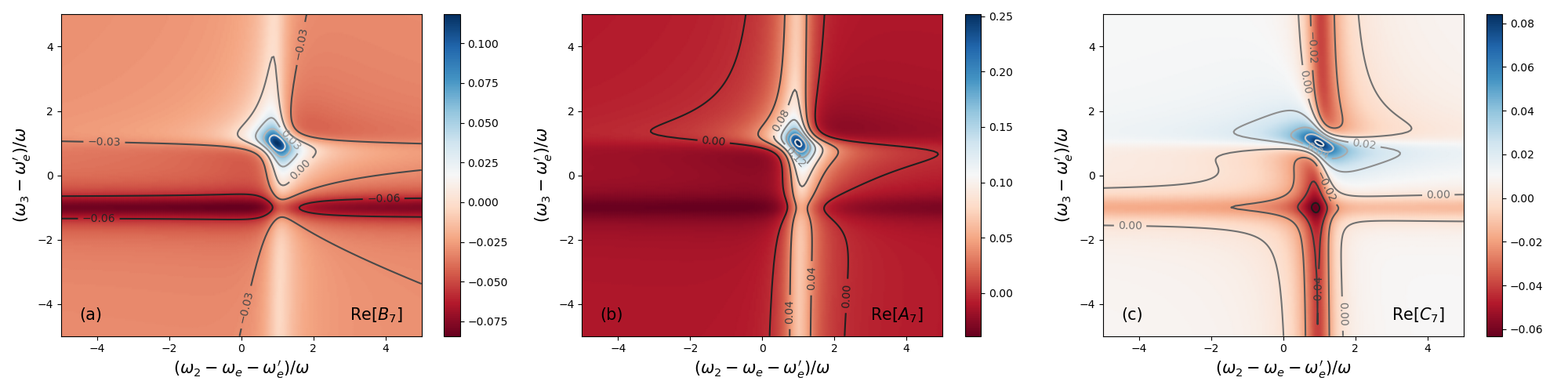}
    \caption{{\color{black}Spectral features associated to the first response function related to double quantum coherence [Fig. \ref{fig3}(g)], Fourier transformed with respect to $t_2$ and $t_3$, for $t_1=0$. The contour plots refer to the terms in the response function that are: quadratic in $z_1$ and of zero-th order in $\hat{\mu}_1$ [$B_7$, Fourier transform of $g^{(e)}_{121}\tilde{B}_7$, see Eq. (\ref{eqB7}), panel (a)]; linear in $z_1$ and $\hat{\mu}_1$ [$A_7$, Fourier transform of $g^{(e)}_{121}\tilde{A}_7$, see Eq. (\ref{eqA7}), panel (b)]; of zero-th order in $z_1$ and quadratic in $\hat{\mu}_1$ [$C_7$, Fourier transform of $g^{(e)}_{121}\tilde{C}_7$, see Eq. (\ref{eqC7}), panel (c)]. The line broadening is determined by $\gamma=0.4\,\omega$, $\hbar\omega_e=\epsilon_1-\epsilon_0$ is the energy difference between the ground and first excited electronic states.}}
    \label{fDQCf}
\end{figure*}

\begin{figure*}
    \centering
    \includegraphics[width=0.95\linewidth]{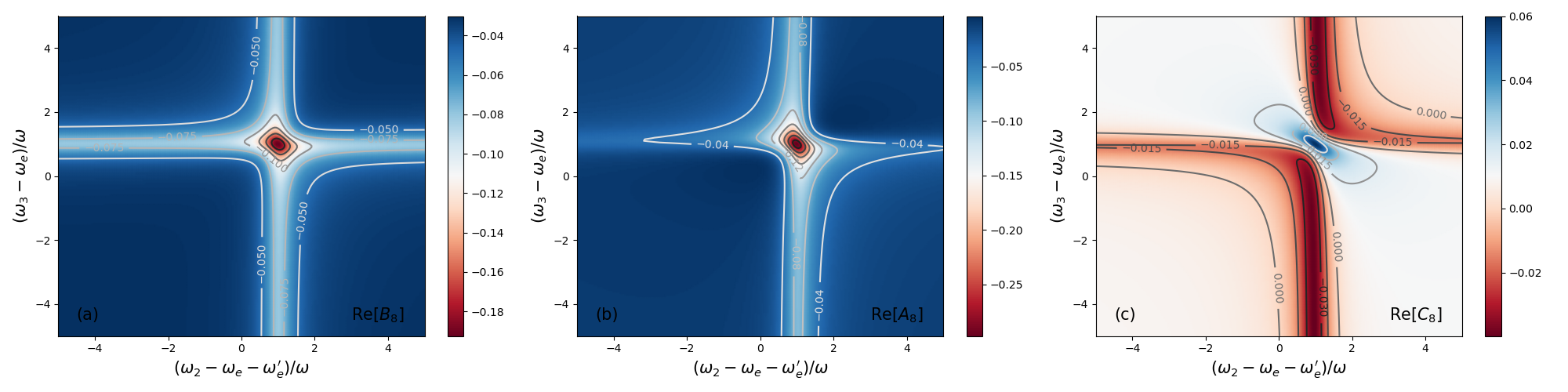}
    \caption{{\color{black}Spectral features associated to the second response function related to double quantum coherence [Fig. \ref{fig3}(h)], Fourier transformed with respect to $t_2$ and $t_3$, for $t_1=0$. The contour plots refer to the terms in the response function that are: quadratic in $z_1$ and of zero-th order in $\hat{\mu}_1$ [$B_8$, Fourier transform of $g^{(e)}_{121}\tilde{B}_8$, see Eq. (\ref{eqB8}), panel (a)]; linear in $z_1$ and $\hat{\mu}_1$ [$A_8$, Fourier transform of $g^{(e)}_{121}\tilde{A}_8$, see Eq. (\ref{eqA8}), panel (b)]; of zero-th order in $z_1$ and quadratic in $\hat{\mu}_1$ [$C_8$, Fourier transform of $g^{(e)}_{121}\tilde{C}_8$, see Eq. (\ref{eqC8}), panel (c)]. The line broadening is determined by $\gamma=0.4\,\omega$, $\hbar\omega_e=\epsilon_1-\epsilon_0$ is the energy difference between the ground and first excited electronic states.}}
    \label{fDQCs}
\end{figure*}

The results presented in this paper are based on the assumption that the non-Condon effects can be accounted for by the second term in the Taylor expansion of the dipole operator with respect to the nuclear coordinate [Eq. (\ref{eq93})]. The generalization of the approach to cases where this approximation is not valid and additional terms in the expansion need to be taken into account is conceptually straightforward. In order to illustrate this, let us consider the third term in the Taylor expansion of the dipole operator (quadratic in the nuclear coordinate), which we rewrite in the normally-ordered form:
\begin{gather}
\hat{\mu}_2 \otimes (\hat{a}^\dagger+\hat{a})^2 = \hat{\mu}_2 \otimes \left[(\hat{a}^\dagger)^2+\hat{a}^2+2\hat{a}^\dagger\hat{a}+\hat{\mathcal{I}}\right]\,. 
\end{gather}
The powers of the annihilation and creation operators give rise to the following contributions:
\begin{gather}
\frac{\langle \alpha | \hat{U}_{\color{black}M+1}\dots \hat{U}_k\,(\hat{a}^\dagger)^m\,\hat{U}_{k-1}\dots \hat{U}_1 |\alpha\rangle}{\langle \alpha | \hat{U}_{\color{black}M+1}\dots \hat{U}_k\,\hat{U}_{k-1}\dots \hat{U}_1 |\alpha\rangle} = \left[c_k^{(M)}\right]^m\label{eqw1}\\
\frac{\langle \alpha | \hat{U}_{\color{black}M+1}\dots \hat{U}_k\,\hat{a}^n\,\hat{U}_{k-1}\dots \hat{U}_1 |\alpha\rangle}{\langle \alpha | \hat{U}_{\color{black}M+1}\dots \hat{U}_k\,\hat{U}_{k-1}\dots \hat{U}_1 |\alpha\rangle} = \left[a_k^{(M)}\right]^n\\
\frac{\langle \alpha | \hat{U}_{\color{black}M+1}\dots \hat{U}_k\,(\hat{a}^\dagger)^p(\hat{a})^q\,\hat{U}_{k-1}\dots \hat{U}_1 |\alpha\rangle}{\langle \alpha | \hat{U}_{\color{black}M+1}\dots \hat{U}_k\,\hat{U}_{k-1}\dots \hat{U}_1 |\alpha\rangle} \!=\! \left[c_k^{(M)}\right]^p\,\left[a_k^{(M)}\right]^q.\label{eqw2}
\end{gather}
One can thus proceed with second- or higher-order powers of the nuclear coordinates along the same lines as for the linear term: $(i)$ writing the power of the position operator $\hat{X}^k$ in terms of the annihilation and creation operators, normally ordered (i.e. creation operators to the left, annihilation to the right); $(ii)$ applying the above Eqs. (\ref{eqw1}-\ref{eqw2}) to each of the terms resulting from the expansion of $\hat{X}^k$; $(iii)$ combining the contributions coming from $\hat{\mu}_2$ with those deriving from $\hat{\mu}_0$ and $\hat{\mu}_1$, along the lines shown in the Appendices B-D.

\subsubsection{Multiple vibrational modes}

The extension to a system with multiple uncoupled vibrational modes is also straightforward. Let us consider the dipole operator
\begin{gather}
    \hat{\mu} = \hat{\mu}_0 + \sum_{\xi=1}^K \hat{\mu}_{1,\xi}\otimes (\hat{a}_\xi^\dagger+\hat{a}_\xi)\,.
\end{gather}
Here, the first term accounts for the FC transitions, and each of the additional terms accounts for transitions that modify the state of one vibrational mode (specified by the index $\xi$), while leaving the other unaffected. One can easily verify that the overall response function is still given by the product of an electronic and a vibrational component. This consists in the sum of $(K+1)^{M+1}$ contributions, corresponding to the different ways in which the $K+1$ terms of the dipole operators can appear in the $M+1$ positions within the propagator. Each of these contributions corresponds to the product of the single-mode propagators that have been derived in the present work.
}

{\color{black}
\subsubsection{Dephasing and relaxation}

The approach developed so far is based on a simulation of the time evolution at the Hamiltonian level (closed quantum system). At a simple phenomenological level, the effect of the environment can also be included. Pure dephasing affecting the electronic degrees of freedom can be accounted for by multiplying the function $g^{(e)}_{\bf e}$ by a function $e^{-\gamma t_k}$ for each of the waiting times $t_k$ where the electronic states in the ket and in the bra are different. Relaxation affecting the electronic degrees of freedom can also be accounted for by multiplying $g^{(e)}_{\bf e}$ by a function $e^{-\Gamma t_k}$ for each of the waiting times $t_k$ where the electronic states in both the ket and the bra are excited (either identical or different from each other), and by a function $e^{-\Gamma t_l/2}$ for each of the waiting times $t_l$ where only one of the electronic states (ket or bra) is excited. This results in the following prefactors for the response functions corresponding to the Feynman diagrams in Fig. \ref{fig3}: (a,b) $e^{-(\gamma+\Gamma/2)(t_1+t_3)}$; (c,d) $e^{-(\gamma+\Gamma/2)(t_1+t_3)-\Gamma t_2}$; (e,f) $e^{-\gamma (t_1+t_3)-\Gamma (t_1/2+t_2+t_3)}$; (g) $e^{-\gamma (t_1+t_2+t_3)-\Gamma (t_1/2+t_2/2+t_3)}$; (h) $e^{-\gamma(t_1+t_2+t_3)-\Gamma(t_1+t_2+t_3)/2}$. The generalization to the case where the relaxation and dephasing rates are state-dependent is straightforward. Relaxation in the vibrational degrees of freedom can be simulated by adding a non-Hermitian term to the Hamiltonian\cite{Quintela22a}. While these exponential prefactors allow one to consistently include within the present approach the effects of dephasing, they only partially account for the effects of relaxation. In particular, the effect of the population loss from the initial electronic state is correctly described, but this is not the case for the population gain in the final electronic state. For example, the relaxation $|1\rangle \rightarrow |0\rangle$ during the second waiting time implies an exponential decay of the response function related to the Feynman diagram in Fig. \ref{fig3}(c), given by the prefactor $e^{-\Gamma t_2}$. However, it also implies an increase in the response function corresponding to the diagram in Fig. \ref{fig3}(a), which cannot easily be included within the present theoretical framework, essentially because the vibrational mode corresponding to the final electronic state would no longer be in a coherent state\cite{Troiani23a}. 
}

\section{Resulting spectral features\label{s4}}

As shown in the previous Section, the HT corrections modify the FC response functions by adding a multiplicative term, which consists of a combination of complex oscillating functions of the waiting times ($\chi_k = e^{-i\omega t_k}$ or $\chi_k^*$). When considering the response functions in the frequency domain, this multiplicative term thus gives rise to a series of replicas of the FC peaks, each displaced by an integer multiple of the vibrational frequency $\omega$. In the semi-impulsive limit, to which we refer in the present Section, the properties of the response functions can be directly related to the spectral features observed in multidimensional spectroscopy\cite{Hamm2011}.

Let us consider, as a representative example, a third-order response function for a three-level system ($N=3$), {\color{black}and a vibrational mode initialized in the ground state ($\alpha=0$, zero-temperature limit). Under the assumption that both the vibrational-mode displacements parameters $z_k$ and the ratio between the matrix elements of $\hat{\mu}_0$ and $\hat{\mu}_1$ are much smaller than 1, we focus on the contributions that are of lowest (but nonzero) order in these quantities. These contributions will be grouped into three terms. The first, denoted with $\tilde{B}$, comes from the FC response function, and specifically from the second term in the Taylor expansion of $g^{(v)}_{ijk}$. This corresponds to $f$, the argument of the exponential [Eq. (\ref{eq91})], which is quadratic in the displacements $z_k$ (and of zero-th order in $\hat{\mu}_1$. The second term ($\tilde{A}$) arises from the HT coupling, and specifically from the function $r_{jkl,1}$, which is linear both in the $z_k$ and in $\hat{\mu}_1$. The third term ($\tilde{C}$) is also related to the HT coupling, and specifically comes from those contributions in $r_{jkl,2}$ that are (quadratic in $\hat{\mu}_1$ and) of the zero-th order in the $z_k$. 

We consider rephasing and non-rephasing response functions corresponding to different processes (Fig. \ref{fig3}): (a,b) ground state bleaching, (c,d) stimulated emission, (e,f) excited state absorption, and (g,h) double quantum coherence. Following a widespread convention, we perform the Fourier transform with respect to the first and third waiting times in the first six cases (for $t_2=0$, and thus $\chi_2=1$) and with respect to the second and third waiting times (for $t_1=0$, and thus $\chi_1=1$) in the last two cases. The Fourier transforms of $g^{(e)}_{jkl}\tilde{A}$, $g^{(e)}_{jkl}\tilde{B}$, and $g^{(e)}_{jkl}\tilde{C}$ are denoted by $A$, $B$, and $C$, respectively. Details on the derivations of the simulated spectra are provided in Appendix D. 

From the equations derived in the previous Section, it follows that the response functions in the presence of HT coupling is given by the product of the FC vibrational response function, $g^{(v)}_{ijk}$, multiplied by combinations of oscillatory functions ($\chi_k$ and $\chi_k^*$). Each of these factors produces a replica of the entire FC spectrum, spectrally displaced by integer multiples of the vibrational frequency $\omega$. The way in which the different contributions combine depends on a number of system-dependent parameters and specifically on the matrix elements of the operators $\hat{\mu}_k$, which determine the coefficients $C_{abcb}^{jkl}$, and on the displacements $z_k$. This makes it difficult in general to identify specific patterns in the 2D spectra that can be unambiguously related to the HT coupling, even if this identification was shown to be possible in some cases\cite{Bizimana17a}. In the following, we set for simplicity $z_1=-z_2=0.1$ and $C_{abcd}^{jkl}$ to 1, 0.1, and 0.01 when the coefficient is of order 0, 1, and 2 in $\hat{\mu}_1$, respectively. Besides, we plot the terms {\color{black}$A$, $ B$, and $C$} separately, for each of the eight response functions specified by the Feynman diagrams in Fig. \ref{fig3}. These include the rephasing and non-rephasing response functions corresponding to ground state bleaching (Figs. \ref{fGSBr}-\ref{fGSBnr}), stimulated emission (Figs.\ref{fSEr}-\ref{fSEnr}), excited state absorption (Figs. \ref{fESAr}-\ref{fESAnr}), and the two contributions related to double quantum coherence (Figs. \ref{fDQCf}-\ref{fDQCs}). The plotted functions represent a contribution to the response functions, typically dominated by the zero-phonon peak (not shown). Given the truncation with respect to the expansion in the displacements $z_j$ and in $\hat\mu_1$ that we consider in these examples, one can only have terms that are linear in $\chi_j$ or $\chi_j^*$, giving rise to secondary peaks displaced by $\pm\omega$ with respect to the main one. Further terms in the expansion, of higher order in $z_j$ and in $\hat\mu_1$, can give rise to replicas of the main peak displaced by $\pm k\omega$, with $k>1$. The peaks appearing, for example, in Fig. \ref{fGSBr}(a), are related to terms (in $\tilde A_1$) like $\chi_1^*\chi_3$; the linear features appearing, for example, in Fig. \ref{fGSBnr}(b) are related to terms (in $\tilde B_2$) like $\chi_1$ (negative vertical feature) and $\chi_3$ (positive horizontal feature). }

{\color{black} In all the above plots the two-dimensional map corresponds to fixed values of a waiting time $t_k$ [$k=2$ in panels (a-f), $k=1$ in (g-h)]. The dynamics of the spectral features is obtained by considering the evolution of the map as a function of $t_k$, and presents features that depend on the system parameters. Some general indication can however be obtained from the expressions of the functions $\tilde A_j$, $\tilde B_j$, and $\tilde C_j$, reported in Appendix D.7. For example, the peak appearing at $\omega_1=-\omega_e-\omega$ and $\omega_3=\omega_e-\omega$ in the rephasing response function related to ground state bleaching, first HT contribution [$A_1$, Fig. \ref{fGSBr}(b)] comes from the terms $-(C_{0010}^{101}+C_{1000}^{101}) \chi_{13}^*$ in Eq. (\ref{eqA1}), being $\chi_{13}^*=e^{i\omega(t_1+t_2+t_3)}$. As a result, the dependence on $t_2$ of this term is given by the exponential function $e^{i\omega t_2}$. The linear feature at $\omega_1=-\omega_e-\omega$ related to the second HT contribution [$C_1$, Fig. \ref{fGSBr}(c)] comes from the terms $C_{1001}^{101} \chi_{12}^* + C_{1100}^{101} \chi_{1}^*$ in Eq. (\ref{eqC1}). Its dependence on $t_2$ is therefore given by $C_{1001}^{101} e^{i\omega t_2} + C_{1100}^{101}$. Similarly, the peak appearing at $\omega_2=\omega_e+\omega_e'+\omega$ and $\omega_3=\omega_e'+\omega$ in the first response function related to double quantum coherence, first HT contribution [$A_7$, Fig. \ref{fDQCf}(b)] comes from the terms $[C_{0100}^{121}z_1-C_{0001}^{121}(z_1+z_2)] \chi_{13}$ in Eq. (\ref{eqA7}), being $\chi_{13}=e^{-i\omega(t_1+t_2+t_3)}$. As a result, the dependence on {\color{black}$t_1$} of this term is given by the exponential function $e^{-i\omega t_1}\,e^{-(\gamma+\Gamma/2+i\omega_e)t_1}$, where the last factor comes from the electronic component of the response function. Finally, the linear feature at $\omega_1=\omega_e+\omega_e'+\omega$ related to the second HT contribution [$C_7$, Fig. \ref{fDQCs}(c)] comes from the terms $C_{1001}^{121} \chi_{12} + C_{0011}^{121} \chi_{1}$ in Eq. (\ref{eqC7}). Its dependence on {\color{black}$t_1$} therefore {\color{black} coincides with that obtained for $A_7$.}}

\section{Conclusions\label{s5}}

In conclusion, we have derived analytical solutions for the response function of the displaced harmonic oscillator model. The method applies to models comprising an arbitrary number $N$ of electronic levels and for arbitrary order $M$ of the light-matter interaction: the solutions are given in a closed form for $M\le 3$ and (simply to avoid lengthy expressions) in a recursive form for $M>3$. The approach is based on the use of a coherent state representation of the vibrational states, and on the reduction of the vibrational propagator entering the expression of the response function to an overlap between a pair of coherent states, multiplied by a combination of complex exponential functions. The presence of HT corrections results in response functions that are given by the FC response functions, multiplied by combinations of complex oscillating functions. For sufficiently short excitation pulses (semi-impulsive limit), such corrections spectrally result in vibrational replicas of the overall FC features.

The expressions of the response functions \FT{are given for the case where the vibrational mode is initialized to a generic coherent state}. Given that the coherent states provide an overcomplete basis for the vibrational mode, this allows one to derive the expression of the response function corresponding to an arbitrary initial state\FT{, including the thermal state (finite temperature)}. Finally, \FT{we show how the present approach can be extended for the case where non-Condon effects \FT{give rise to} nonlinear terms in the expansion of the transition dipole as a function of nuclear coordinates, and for models that include multiple vibrational modes (in the absence of Duschinsky mode mixing)}.

{\color{black} A Python code for the analytical calculation of the above response functions and for the simulation of the multidimensional spectra is available on GitHub\cite{TroianiGitHub}.}

\appendix 

\acknowledgements 

The author acknowledges useful discussions with Frank Ernesto Quintela Rodriguez.

\section{Commutation relations and notation}

We start by recalling the commutation relations between the annihilation and creation operators on the one hand, and the displacement, rotation, and time-evolution operators on the other hand. 

The relevant commutation relations involving the displacement operator $\hat{\mathcal{D}}(z)=\exp [z(\hat{a}^\dagger - \hat{a})]$, with $z$ a real number, are given by the following equations:
\begin{gather}
    \hat{a}\, \hat{\mathcal{D}} (z)  = \hat{\mathcal{D}} (z)\, (\hat{a} + z) \\
    \hat{\mathcal{D}} (z)\, \hat{a}^\dagger = (\hat{a}^\dagger -z)\,\hat{\mathcal{D}} (z) \,.
\end{gather}

The relevant commutation relations involving the rotation operator $\hat{R}(\phi)=\exp (i\phi \hat{a}^\dagger \hat{a})$, with $\phi$ a real number, are given by the following equations:
\begin{gather}
\hat{a}\,\hat{R}(\phi)=\hat{R}(\phi)\, \hat{a}\,e^{i\phi}\\
\hat{R}(\phi)\, \hat{a}^\dagger = \hat{a}^\dagger\, \hat{R}(\phi)\, e^{i\phi} .
\end{gather}

The time evolution operator $\hat{U}$ corresponds to a product of the displacement and rotation operators:
\begin{gather}
 \hat{U} (t,z) = \hat{\mathcal{D}} (-z)\, \hat{R} (-i\omega t)\, \hat{\mathcal{D}} (z) .   
\end{gather}
From this and from the above equations, it follows that:
\begin{gather}
    \hat{a}\,\hat{U} = \hat{U}\, [\hat{a}\,e^{-i\omega t} + z\,(e^{-i\omega t}-1)] \label{eq04} \\ \label{eq02}
    \hat{U}\,\hat{a}^\dagger = [\hat{a}^\dagger\,e^{-i\omega t} - z\,(e^{-i\omega t}-1)] \, \hat{U}\,.
\end{gather}

In order to simplify the expressions of the $M$-th order response functions reported hereafter, we introduce a compact notation. In particular, we introduce the oscillating terms:
\begin{gather}\label{eq12}
\chi_{kl} \equiv e^{-i\omega \sum_{j=k}^l t_j}\,,\ \chi_k \equiv \chi_{kk} = e^{-i\omega t_k} \,, 
\end{gather}
where $t_k$ ($k = 1,\dots,M$) are the waiting times.
The above time-dependent functions are expressed as follows: 
\begin{gather}\label{eq11}
f_{k,l} \equiv z_{e_k}(\chi_l-1) \,,
\end{gather}
where and $|e_k\rangle$ is the electronic state that determines the displacement $z_{e_k}$ of the vibrational mode. 
Along the same lines, the following compact notation is introduced for the time-evolution operators:
\begin{gather}
\hat{U}_k \equiv \hat{U} (t_k,z_{e_k}) \,,\  
\hat{U}_{M+1} \equiv \hat{U} \left(-\sum_{k=1}^{M} t_k,0\right) .
\end{gather}

\FT{We finally remind that a coherent state of a quantum harmonic oscillator, specified by the complex number $\alpha$, is expressed as follows in terms of the Fock states $|n\rangle$:
\begin{equation}
|\alpha\rangle = e^{-|\alpha|^2/2} \sum_{n=0}^\infty \frac{\alpha^n}{\sqrt{n!}}|n\rangle .
\end{equation}
From this it follows that $\hat{a}|\alpha\rangle = \alpha |\alpha\rangle$ and $\langle\alpha|\hat{a}^\dagger = \langle\alpha| \alpha^*$, two equations that are used hereafter in the derivation of the HT contributions to the response functions.}

\section{First-order response function}

The different terms that contribute to the vibrational component of the first-order response function are divided hereafter according to the order in which the FC and the HT components appear and, for each order, to the amplitude of the nonlinear transition, determined by the electronic degrees of freedom.

\subsection{FC term}
The FC term is of order 2 and 0 in $\hat{\mu}_0$ and $\hat{\mu}_1$, respectively. The vibrational component of the response function corresponding to the electronic state $|j\rangle$ reads\cite{Quintela22a}: 
\begin{gather}
    g_j^{(v)} = \exp\left[ z_j^2 \left( \chi_1 - 1\right) \right]\,,
\end{gather}
where \FT{$g_j^{(v,\alpha)}\equiv \langle \alpha | \,\hat{U}_2\, \hat{U}_1 | \alpha \rangle$, $g_j^{(v)}\equiv g_j^{(v,0)}$.} \FT{The derivation of the expressions corresponding to $\alpha\neq 0$ is discussed in Ref. \onlinecite{Quintela22a}.}

In the general expression of the response function, the above vibrational terms are multiplied by the electronic contribution $C^{j}_{00}g_j^{(e)}$, where
\begin{gather}\label{eq14}
    C^{j}_{ab} = \langle 0 | \hat{\mu}_a | j \rangle\,\langle j | \hat{\mu}_b | 0 \rangle\,.
\end{gather}

\subsection{{\color{black}Linear} HT terms}
The first group of mixed terms we consider is of order 1 both in $\hat{\mu}_0$ and in $\hat{\mu}_1$. The terms resulting from the presence of the annihilation and creation operators in the first position (to the right of $\hat{U}_1$) are given by: 
\begin{gather}
    f^j_{1|}\equiv \langle \FT{\alpha} | \,\FT{\hat{U}_2}\,\hat{U}_1\, \hat{a} \,| \FT{\alpha} \rangle / g_j^{(v\FT{,\alpha})} \equiv a^{(1)}_1 = \FT{\alpha} \label{eq01p} \\ \label{eq01}
    f^j_{|1}\equiv \langle \FT{\alpha} | \,\FT{\hat{U}_2}\,\hat{U}_1\, \hat{a}^\dagger \,| \FT{\alpha} \rangle / g_j^{(v\FT{,\alpha})} \equiv c^{(1)}_1 =
     \FT{\alpha^*}-f_{j,1}\,.
\end{gather}
The above equations are obtained by applying Eqs. (\ref{eq04}-\ref{eq02}); the definition of $f_{j,1}$ is given in Eq. (\ref{eq11}). In the notation adopted for the functions $f^{j\dots}_{k\dots|l\dots}$, the superscript specifies the electronic state(s), while the numbers on the left- and right-hand sides of the vertical bar in the subscript specify the positions of the annihilation and creation operators, respectively. In the general expression of the response function the above vibrational terms are multiplied by the electronic contribution $C^{j}_{01}\,g_j^{(e)}$.

The terms resulting from the presence of the annihilation and creation operators in the second position (to the left of $\hat{U}_1$) are given by the equations: 
\begin{gather}
    f^j_{2|}\equiv \langle \FT{\alpha} | \, \FT{\hat{U}_2}\, \hat{a}\, \hat{U}_1\, | \FT{\alpha} \rangle / g_j^{(v\FT{,\alpha})} \equiv a^{(1)}_2 = 
    f_{j,1} \FT{+\alpha \chi_1} \\ \label{eq03}
    f^j_{|2}\equiv \langle \FT{\alpha} | \, \FT{\hat{U}_2}\, \hat{a}^\dagger\, \hat{U}_1\, | \FT{\alpha} \rangle / g_j^{(v\FT{,\alpha})} \equiv c^{(1)}_2 = \FT{\alpha^*\chi_1^* }\,,
\end{gather}
which are also based on Eqs. (\ref{eq04}-\ref{eq02}) and on the definition in Eq. (\ref{eq11}).
In the general expression of the response function, the above vibrational terms are multiplied by the electronic contribution $C^{j}_{10}\,g_j^{(e)}$.

\subsection{{\color{black}Quadratic} HT terms}
The HT terms are of the order 0 and 2 in $\hat{\mu}_0$ and $\hat{\mu}_1$, respectively. The terms resulting from the presence of the annihilation and creation operators in the three positions are given by the following: 
\begin{gather}
    f_{12|}^j\equiv\langle \FT{\alpha} | \, \FT{\hat{U}_2}\,\hat{a}\, \hat{U}_1\, \hat{a} \,| \FT{\alpha} \rangle / g_j^{(v\FT{,\alpha})} = 
    a^{(1)}_2\,a^{(1)}_1  \label{eq07}\\
    f_{1|2}^j\equiv\langle \FT{\alpha} | \, \FT{\hat{U}_2}\,\hat{a}^\dagger\, \hat{U}_1\, \hat{a} \,| \FT{\alpha} \rangle / g_j^{(v\FT{,\alpha})} = 
    c^{(1)}_2\,a^{(1)}_1  \\
    f_{|12}^j\equiv\langle \FT{\alpha} | \, \FT{\hat{U}_2}\,\hat{a}^\dagger\, \hat{U}_1\, \hat{a}^\dagger \,| \FT{\alpha} \rangle / g_j^{(v\FT{,\alpha})} = 
    c^{(1)}_2\,c^{(1)}_1  \label{eq08} \\
    f_{2|1}^j\equiv\langle \FT{\alpha} | \,\FT{\hat{U}_2}\, \hat{a}\, \hat{U}_1\, \hat{a}^\dagger \,| \FT{\alpha} \rangle / g_j^{(v\FT{,\alpha})} = 
    \chi_1 + a_2^{(1)} c_1^{(1)} . \label{eq09}
\end{gather}
In the general expression of the response function, the above vibrational terms are multiplied by the electronic contribution $C^{j}_{11}\,g_j^{(e)}$.

\section{Second-order response function}

The different terms that contribute to the vibrational component of the second-order response function are divided hereafter according to the order in which the FC and the HT components appear and, for each order, according to the amplitude of the nonlinear transition determined by the electronic degrees of freedom.

\subsection{FC term}
The FC term is of order 3 and 0 in $\hat{\mu}_0$ and $\hat{\mu}_1$, respectively. The vibrational component of the response function corresponding to the sequence of electronic states $|j\rangle \rightarrow |k\rangle$ is given by\cite{Quintela22a}: 
\begin{gather}
    g_{jk}^{(v)} 
    = \exp\left[z_jz_{jk} (\chi_1\!-\!1)\!+\!z_{kj}z_k(\chi_2\!-\!1)\!+\!z_jz_k(\chi_{12}\!-\!1)\right]
\end{gather}
where \FT{$g_{jk}^{(v,\alpha)}\equiv \langle \alpha | \,\hat{U}_3\,\hat{U}_2\, \hat{U}_1 | \alpha \rangle$, $g_{jk}^{(v)}\equiv g_{jk}^{(v,0)}$,} $z_{jk}\equiv z_j-z_k$, and the $\chi_k$ are given in Eq. (\ref{eq12}). \FT{The derivation of the expressions corresponding to $\alpha\neq 0$ is discussed in Ref. \onlinecite{Quintela22a}.}

In the general expression of the response function, the above vibrational terms are multiplied by the electronic contribution $C^{jk}_{000}g_{jk}^{(e)}$, where
\begin{gather}\label{eq19}
    C^{jk}_{abc} = \langle 0 | \hat{\mu}_a | k \rangle\,\langle k | \hat{\mu}_b | j \rangle\,\langle j | \hat{\mu}_c | 0 \rangle\,.
\end{gather}

\subsection{{\color{black}Linear} HT terms}
The first group of mixed terms we consider is of order 2 and 1 in $\hat{\mu}_0$ and $\hat{\mu}_1$, respectively. In particular, the nonzero terms resulting from the presence of the annihilation and creation operators in the first position (to the right of $\hat{U}_1$) are given by the following: 
\begin{gather}\label{eq27}
    f^{jk}_{|1}\equiv\langle \FT{\alpha} | \, \hat{U}_3\, \hat{U}_2\, \hat{U}_1\, \hat{a}^\dagger\, | \FT{\alpha} \rangle / g_{jk}^{(v\FT{,\alpha})} \equiv c^{(2)}_1 \!=\! \FT{\alpha^*}
    \!-\!f_{j,1} \!-\! \chi_1 f_{k,2} \\
        {\color{black}f^{jk}_{1|}\equiv\langle \alpha | \, \hat{U}_3\, \hat{U}_2\, \hat{U}_1\, \hat{a}\, | \alpha \rangle / g_{jk}^{(v\FT{,\alpha})} \equiv a^{(2)}_1 = \alpha} .
\end{gather}
The definitions of $f_{i,k}$ and $\chi_k$ are given in Eq. (\ref{eq11}) and in Eq. (\ref{eq12}), respectively. 
In the general expression of the response function, the above vibrational terms are multiplied by the electronic contribution $C^{jk}_{001}\,g_{jk}^{(e)}$.

The nonzero terms resulting from the presence of the annihilation and creation operators in the second position (to the right of $\hat{U}_2$) are given by:
\begin{gather}\label{eq28}
    f^{jk}_{2|}\equiv\langle \FT{\alpha} | \, \FT{\hat{U}_3}\, \hat{U}_2\, \hat{a}\, \hat{U}_1\, | \FT{\alpha} \rangle / g_{jk}^{(v\FT{,\alpha})} \equiv a^{(2)}_2 = 
    f_{j,1} \FT{+\chi_1\alpha}\\
    f^{jk}_{|2}\equiv\langle \FT{\alpha} | \, \FT{\hat{U}_3}\, \hat{U}_2\, \hat{a}^\dagger\, \hat{U}_1\, | \FT{\alpha} \rangle / g_{jk}^{(v\FT{,\alpha})} \equiv c^{(2)}_2 = 
    \FT{\chi_1^*\alpha^*}-f_{k,2}\,. \label{eq29}
\end{gather}
In the general expression of the response function, the above vibrational terms are multiplied by the electronic contribution $C^{jk}_{010}g_{jk}^{(e)}$.

The nonzero term resulting from the presence of the annihilation and creation operators in the third position (to the left of $\hat{U}_2$) is given by:
\begin{gather}\label{eq30}
    f_{3|}^{jk}\!\equiv\!\langle \FT{\alpha} | \, \FT{\hat{U}_3}\,\hat{a}\, \hat{U}_2\, \hat{U}_1\, | \FT{\alpha} \rangle / g_{jk}^{(v\FT{,\alpha})} \!\equiv\! a^{(2)}_3 \!=\! 
    f_{k,2} \!+\! \chi_2 f_{j,1}\FT{+\chi_{12}\alpha}\\
        {\color{black}f_{|3}^{jk}\equiv\langle \alpha | \, \FT{\hat{U}_3}\,\hat{a}^\dagger\, \hat{U}_2\, \hat{U}_1\, | \alpha \rangle / g_{jk}^{(v\FT{,\alpha})} \equiv c^{(2)}_3 = 
    \chi^*_{12} \alpha^*}\,.
\end{gather}
In the general expression of the response function, the above vibrational terms are multiplied by the electronic contribution $C^{jk}_{100}g_{jk}^{(e)}$.

\subsection{{\color{black}Quadratic} HT terms}
The \FT{terms that are nonzero for $\alpha=0$} resulting from the presence of the annihilation and creation operators in the first and second positions are given by: 
\begin{gather}\label{eq31}
    f^{jk}_{|12}\equiv\langle \FT{\alpha} | \, \FT{\hat{U}_3}\, \hat{U}_2\, \hat{a}^\dagger\, \hat{U}_1\, \hat{a}^\dagger\, | \FT{\alpha} \rangle / g_{jk}^{(v\FT{,\alpha})} = 
    c_1^{(2)} c_2^{(2)} \\ \label{eq32}
    f^{jk}_{2|1}\equiv\langle \FT{\alpha} | \, \FT{\hat{U}_3}\, \hat{U}_2\, \hat{a}\, \hat{U}_1\, \hat{a}^\dagger\, | \FT{\alpha} \rangle  / g_{jk}^{(v\FT{,\alpha})} 
    = \chi_1+c_1^{(2)} a_2^{(2)} \,.
\end{gather}
In the general expression of the response function, the above vibrational terms are multiplied by the electronic contribution $C^{jk}_{011}\,g_{jk}^{(e)}$.

The \FT{terms that are nonzero for $\alpha=0$} resulting from the presence of the annihilation and creation operators in the second and third positions are given by: 
\begin{gather}\label{eq33}
    f^{jk}_{23|}\equiv\langle \FT{\alpha} | \, \FT{\hat{U}_3}\, \hat{a}\, \hat{U}_2\, \hat{a}\, \hat{U}_1\, | \FT{\alpha} \rangle / g_{jk}^{(v\FT{,\alpha})} = a^{(2)}_2 a^{(2)}_3 \\
    f^{jk}_{3|2}\equiv\langle \FT{\alpha} | \, \FT{\hat{U}_3}\, \hat{a}\,\hat{U}_2\, \hat{a}^\dagger\, \hat{U}_1\, | \FT{\alpha} \rangle / g_{jk}^{(v\FT{,\alpha})} 
    = \chi_2 + c^{(2)}_2 a^{(2)}_3\,. \label{eq34}
\end{gather}
In the general expression of the response function, the above vibrational terms are multiplied by the electronic contribution $C^{jk}_{110}\,g_{jk}^{(e)}$.

The \FT{term that is nonzero for $\alpha=0$} resulting from the presence of the annihilation and creation operators in the first and third positions are given by: 
\begin{gather}
    f_{3|1}^{jk}\equiv\langle \FT{\alpha} | \,\FT{\hat{U}_3}\, \hat{a}\,\hat{U}_2\, \hat{U}_1\, \hat{a}^\dagger\, | \FT{\alpha} \rangle  / g_{jk}^{(v\FT{,\alpha})} 
    = \chi_{12} +a_3^{(2)} c_1^{(2)}\label{eq35}\,.
\end{gather}
In the general expression of the response function, the above vibrational terms are multiplied by the electronic contribution $C^{jk}_{101}\,g_{jk}^{(e)}$.

\subsection{{\color{black}Cubic} HT terms}
The HT terms are of the order 0 and 3 in $\hat{\mu}_0$ and $\hat{\mu}_1$, respectively. The \FT{terms that are nonzero for $\alpha=0$} resulting from the presence of the annihilation and creation operators in the four positions are given by: 
\begin{gather}
    f^{jk}_{23|1}\equiv\langle \FT{\alpha} | \,\FT{\hat{U}_3}\, \hat{a}\, \hat{U}_2\, \hat{a}\, \hat{U}_1\, \hat{a}^\dagger\,| \FT{\alpha} \rangle / g_{jk}^{(v\FT{,\alpha})} = \nonumber\\ {\color{black} a_3^{(2)} a_2^{(2)} c_1^{(2)}+\chi_1\,a_3^{(2)}+\chi_{12}\,a_2^{(2)}}\label{eq36}\\ 
    f^{jk}_{3|12}\equiv\langle \FT{\alpha} | \,\FT{\hat{U}_3}\, \hat{a}\, \hat{U}_2\, \hat{a}^\dagger\, \hat{U}_1\, \hat{a}^\dagger\,| \FT{\alpha} \rangle / g_{jk}^{(v\FT{,\alpha})} = \nonumber\\ {\color{black} a_3^{(2)} c_2^{(2)} c_1^{(2)}+\chi_2 \,c_1^{(2)}+\chi_{12}\,c_2^{(2)}}\label{eq37}\,.
\end{gather}
In the general expression of the response function, the above vibrational terms are multiplied by the electronic contribution $C^{jk}_{111}\,g_{jk}^{(e)}$.

\section{Third-order response functions\label{ggg}}

The different terms that contribute to the vibrational component of the third-order response function are divided hereafter according to the order in which the FC and the HT components appear and, for each order, to the amplitude of the nonlinear transition, determined by the electronic degrees of freedom.

\subsection{FC term}
The FC term is of order 4 and 0 in $\hat{\mu}_0$ and $\hat{\mu}_1$, respectively. The vibrational component of the response function corresponding to the sequence of electronic states $|j\rangle \rightarrow |k\rangle \rightarrow |l\rangle$ is given by\cite{Quintela22a}: 
\begin{gather}
    g_{jkl}^{(v)} = \exp\left[z_{j}z_{jk}(\chi_1-1)\right.\nonumber\\ +z_{kj}z_{kl}(\chi_2-1)+z_{lk}z_{l}(\chi_3-1)+z_{j}z_{kl}(\chi_{12}-1)\nonumber\\ \left. +z_{kj}z_{l}(\chi_{23}-1)+z_{j}z_{l}(\chi_{13}-1) \right]
\end{gather}
where \FT{$g_{jkl}^{(v,\alpha)}\equiv \langle \alpha | \,\hat{U}_4\,\hat{U}_3\,\hat{U}_2\, \hat{U}_1 | \alpha \rangle$, $g_{jkl}^{(v)}\equiv g_{jkl}^{(v,0)}$}, and the $\chi_k$ are defined in Eq. (\ref{eq12}). \FT{The derivation of the expressions corresponding to $\alpha\neq 0$ is discussed in Ref. \onlinecite{Quintela22a}.} 

In the general expression of the response function, the above vibrational terms are multiplied by the electronic contribution $C^{jkl}_{1234}\,g_{jkl}^{(e)}$, where
\begin{gather}\label{eq38}
    C^{jkl}_{abcd} \!=\! \langle 0 | \hat{\mu}_a | l \rangle\langle l | \hat{\mu}_b | k \rangle\langle k | \hat{\mu}_c | j \rangle\langle j | \hat{\mu}_d | 0 \rangle\,.
\end{gather}

\subsection{{\color{black}Linear} HT terms\label{appsubsec:torf}}
The first group of mixed terms we consider is of order 3 and 1 in $\hat{\mu}_0$ and $\hat{\mu}_1$, respectively. In particular, the terms resulting from the presence of the annihilation and creation operators in the first position (to the right of $\hat{U}_1$) is given by the following: 
\begin{gather}
    \FT{f_{1|}^{jkl}\equiv\langle \alpha | \,\hat{U}_4\, \hat{U}_3\,\hat{U}_2\,\hat{U}_1\, \hat{a}\, | \FT{\alpha} \rangle / g_{jkl}^{(v\FT{,\alpha})} \equiv a^{(3)}_1 =\alpha}\\
    f_{|1}^{jkl}\equiv\langle \FT{\alpha} | \,\FT{\hat{U}_4}\, \hat{U}_3\,\hat{U}_2\,\hat{U}_1\, \hat{a}^\dagger\, | \FT{\alpha} \rangle / g_{jkl}^{(v\FT{,\alpha})} \equiv c^{(3)}_1 \nonumber\\ =\FT{\alpha^*}-(f_{l,3} \chi_{2} + f_{k,2}) \chi_{1} - f_{j,1}\,\,.\label{eq54}
\end{gather}
In the general expression of the response function, the above vibrational terms are multiplied by the electronic contribution $C^{jkl}_{0001}\,g_{jkl}^{(e)}$.

The terms resulting from the presence of the annihilation and creation operators in the second position (to the right of $\hat{U}_2$) are given by:
\begin{gather}\label{eq55}
    f^{jkl}_{2|}\equiv\langle \FT{\alpha} | \,\FT{\hat{U}_4}\, \hat{U}_3\,\hat{U}_2\,\hat{a}\,\hat{U}_1\, | \FT{\alpha} \rangle / g_{jkl}^{(v\FT{,\alpha})} \!\equiv\! a^{(3)}_2 \!=\! f_{j,1} \!+\! \chi_1\, \alpha\\
    f^{jkl}_{|2}\equiv\langle \FT{\alpha} | \,\FT{\hat{U}_4}\, \hat{U}_3\,\hat{U}_2\,\hat{a}^\dagger\,\hat{U}_1\, | \FT{\alpha} \rangle / g_{jkl}^{(v\FT{,\alpha})} \!\equiv\! c^{(3)}_2 \nonumber\\= -\!f_{l,3} \chi_{2}\!-\!f_{k,2} \FT{+\chi_1^*\alpha^*}\,.\label{eq56}
\end{gather}
In the general expression of the response function, the above vibrational terms are multiplied by the electronic contribution $C^{jkl}_{0010}\,g_{jkl}^{(e)}$.

The terms resulting from the presence of the annihilation and creation operators in the third position (to the right of $\hat{U}_3$) are given by:
\begin{gather}\label{eq57}
    f^{jkl}_{3|}\equiv\langle \FT{\alpha} | \,\FT{\hat{U}_4}\, \hat{U}_3\,\hat{a}\,\hat{U}_2\,\hat{U}_1\, | \FT{\alpha} \rangle / g_{jkl}^{(v\FT{,\alpha})} \equiv a^{(3)}_3\nonumber\\= f_{j,1} \chi_{2}\! +\! f_{k,2} \FT{+\chi_{12}\alpha}\\
    f^{jkl}_{|3}\equiv\langle \FT{\alpha} |\,\FT{\hat{U}_4}\, \hat{U}_3\,\hat{a}^\dagger\,\hat{U}_2\,\hat{U}_1\, | \FT{\alpha} \rangle / g_{jkl}^{(v\FT{,\alpha})} \!\equiv\! c^{(3)}_3\! =\! \FT{\chi_{12}^*\alpha^*}\!-\! f_{l,3} \,.\label{eq58}
\end{gather}
In the general expression of the response function, the above vibrational terms are multiplied by the electronic contribution $C^{jkl}_{0100}\,g_{jkl}^{(e)}$.

The terms resulting from the presence of the annihilation and creation operators in the fourth position (to the left of $\hat{U}_3$) is given by:
\begin{gather}
    \FT{f_{|4}^{jkl}\equiv\langle \FT{\alpha} | \,\FT{\hat{U}_4}\, \hat{a}^\dagger\,\hat{U}_3\,\hat{U}_2\,\hat{U}_1\, | \FT{\alpha} \rangle / g_{jkl}^{(v\FT{,\alpha})} \!\equiv\! c^{(3)}_4 = \chi_{13}^*\alpha^*}\\
    f_{4|}^{jkl}\equiv\langle \FT{\alpha} | \,\FT{\hat{U}_4}\, \hat{a}\,\hat{U}_3\,\hat{U}_2\,\hat{U}_1\, | \FT{\alpha} \rangle / g_{jkl}^{(v\FT{,\alpha})} \!\equiv\! a^{(3)}_4 \nonumber\\= (f_{j,1} \chi_{2} \!+\! f_{k,2}) \chi_{3} \!+\! f_{l,3} +\FT{\chi_{13}\alpha}\,. \label{eq59}
\end{gather}
In the general expression of the response function, the above vibrational terms are multiplied by the electronic contribution $C_{1000}^{jkl}\,g_{jkl}^{(e)}$.

The above equations define the functions $a^{(3)}_k$ and $c^{(3)}_k$, which enter the expressions of the terms given below.

\subsection{{\color{black}Quadratic} HT terms\label{appsubsec:torf2}}
The second group of mixed terms we consider is of the second order in both $\hat{\mu}_0$ and $\hat{\mu}_1$. In particular, the \FT{terms that are nonzero for $\alpha=0$} resulting from the presence of the annihilation and creation operators in the first and second positions are given by: 
\begin{gather}\label{eq60}
f^{jkl}_{2|1}\equiv\langle \FT{\alpha} | \,\FT{\hat{U}_4}\, \hat{U}_3\,\hat{U}_2\,\hat{a}\,\hat{U}_1\,\hat{a}^\dagger\, | \FT{\alpha} \rangle / g_{jkl}^{(v\FT{,\alpha})} = \chi_{1}+a^{(3)}_2\,c^{(3)}_1 \\
f^{jkl}_{|12}\equiv\langle \FT{\alpha} | \,\FT{\hat{U}_4}\, \hat{U}_3\,\hat{U}_2\,\hat{a}^\dagger\,\hat{U}_1\, \hat{a}^\dagger\,| \FT{\alpha} \rangle / g_{jkl}^{(v\FT{,\alpha})} = c^{(3)}_2\,c^{(3)}_1 \,.\label{eq61}
\end{gather}
In the general expression of the response function, the above vibrational terms are multiplied by the electronic contribution $C^{jkl}_{0011}\,g_{jkl}^{(e)}$.

The \FT{terms that are nonzero for $\alpha=0$} resulting from the presence of the annihilation and creation operators in the first and third positions are given by:
\begin{gather}\label{eq62}
f^{jkl}_{|13}\equiv\langle \FT{\alpha} | \,\FT{\hat{U}_4}\, \hat{U}_3\,\hat{a}^\dagger\,\hat{U}_2\,\hat{U}_1\,\hat{a}^\dagger\, | \FT{\alpha} \rangle / g_{jkl}^{(v\FT{,\alpha})} = c^{(3)}_3\,c^{(3)}_1 \\
f^{jkl}_{3|1}\equiv\langle \FT{\alpha} | \,\FT{\hat{U}_4}\, \hat{U}_3\,\hat{a}\,\hat{U}_2\,\hat{U}_1\,\hat{a}^\dagger\, | \FT{\alpha} \rangle / g_{jkl}^{(v\FT{,\alpha})} = \chi_{12}+a^{(3)}_3\,c^{(3)}_1\,. \label{eq63}
\end{gather}
In the general expression of the response function, the above vibrational terms are multiplied by the electronic contribution $C^{jkl}_{0101}\,g_{jkl}^{(e)}$.

The \FT{only term that is nonzero for $\alpha=0$} resulting from the presence of the annihilation and creation operators in the first and fourth positions is given by:
\begin{gather}
f_{4|1}^{jkl}\equiv\langle \FT{\alpha} | \,\FT{\hat{U}_4}\,\hat{a}\, \hat{U}_3\,\hat{U}_2\,\hat{U}_1\,\hat{a}^\dagger\, | \FT{\alpha} \rangle / g_{jkl}^{(v\FT{,\alpha})} = \chi_{13}+a^{(3)}_4\,c^{(3)}_1\,. \label{eq64}
\end{gather}
In the general expression of the response function, the above vibrational terms are multiplied by the electronic contribution $C^{jkl}_{1001}g_{jkl}^{(e)}$.

The \FT{terms that are nonzero for $\alpha=0$} resulting from the presence of the annihilation and creation operators in the second and third positions are given by:
\begin{gather}
f^{jkl}_{3|2}\equiv\langle \FT{\alpha} | \,\FT{\hat{U}_4}\,\hat{U}_3\,\hat{a}\,\hat{U}_2\,\hat{a}^\dagger\,\hat{U}_1\, | 0\FT{\alpha}\rangle / g_{jkl}^{(v\FT{,\alpha})} \!=\! \chi_2\!+\! a^{(3)}_3\,c^{(3)}_2 \label{eq65}\\
f^{jkl}_{23|}\equiv\langle \FT{\alpha} | \,\FT{\hat{U}_4}\, \hat{U}_3\,\hat{a}\,\hat{U}_2\,\hat{a}\,\hat{U}_1\, | \FT{\alpha} \rangle / g_{jkl}^{(v\FT{,\alpha})} = a^{(3)}_3\,a^{(3)}_2 \label{eq66}\\
f^{jkl}_{|23}\equiv\langle \FT{\alpha} | \,\FT{\hat{U}_4}\, \hat{U}_3\,\hat{a}^\dagger\,\hat{U}_2\,\hat{a}^\dagger\,\hat{U}_1\, | \FT{\alpha} \rangle / g_{jkl}^{(v\FT{,\alpha})} = c^{(3)}_3\,c^{(3)}_2 \label{eq67}\\
f^{jkl}_{2|3}\equiv\langle \FT{\alpha} | \,\FT{\hat{U}_4}\, \hat{U}_3\,\hat{a}^\dagger\,\hat{U}_2\,\hat{a}\,\hat{U}_1\, | \FT{\alpha} \rangle / g_{jkl}^{(v\FT{,\alpha})} = c^{(3)}_3\,a^{(3)}_2 \,.\label{eq68}
\end{gather}
In the general expression of the response function, the above vibrational terms are multiplied by the electronic contribution $C^{jkl}_{0110}\,g_{jkl}^{(e)}$.

The \FT{terms that are nonzero for $\alpha=0$} resulting from the presence of the annihilation and creation operators in the second and fourth positions are given by:
\begin{gather}
f^{jkl}_{24|}\equiv\langle \FT{\alpha} | \,\FT{\hat{U}_4}\, \hat{a}\,\hat{U}_3\,\hat{U}_2\,\hat{a}\,\hat{U}_1\, | \FT{\alpha} \rangle / g_{jkl}^{(v\FT{,\alpha})} = a^{(3)}_4\,a^{(3)}_2 \label{eq69}\\
f^{jkl}_{4|2}\equiv\langle \FT{\alpha} | \,\FT{\hat{U}_4}\,\hat{a}\, \hat{U}_3\,\hat{U}_2\,\hat{a}^\dagger\,\hat{U}_1\, | \FT{\alpha} \rangle / g_{jkl}^{(v\FT{,\alpha})} = \chi_{23}+a^{\color{black}(3)}_4\,c^{(3)}_2\,. \label{eq70}
\end{gather}
In the general expression of the response function, the above vibrational terms are multiplied by the electronic contribution $C^{jkl}_{1010}g_{jkl}^{(e)}$.

Finally, the \FT{terms that are nonzero for $\alpha=0$} resulting from the presence of the annihilation and creation operators in the third and fourth positions are given by:
\begin{gather}
f^{jkl}_{34|}\equiv\langle \FT{\alpha} | \,\FT{\hat{U}_4}\, \hat{a}\,\hat{U}_3\,\hat{a}\,\hat{U}_2\,\hat{U}_1\, | \FT{\alpha} \rangle / g_{jkl}^{(v\FT{,\alpha})} = a^{(3)}_4\,a^{(3)}_3 \label{eq71}\\
f^{jkl}_{4|3}\equiv\langle \FT{\alpha} | \,\FT{\hat{U}_4}\,\hat{a}\, \hat{U}_3\,\hat{a}^\dagger\,\hat{U}_2\,\hat{U}_1\, | \FT{\alpha} \rangle / g_{jkl}^{(v\FT{,\alpha})} = \chi_{3}+a^{\color{black}(3)}_4\,c^{(3)}_3 \,.\label{eq72}
\end{gather}
In the general expression of the response function, the above vibrational terms are multiplied by the electronic contribution $C^{jkl}_{1100}\,g_{jkl}^{(e)}$.

\subsection{{\color{black}Cubic} HT terms}
The first group of mixed terms we consider is of order 1 and 3 in $\hat{\mu}_0$ and $\hat{\mu}_1$, respectively. In particular, the \FT{terms that are nonzero for $\alpha=0$} resulting from the presence of the annihilation and creation operators are in all positions but the first one are given by: 
\begin{gather}
f^{jkl}_{234|}\equiv\langle \FT{\alpha} | \,\FT{\hat{U}_4}\, \hat{a}\,\hat{U}_3\,\hat{a}\,\hat{U}_2\,\hat{a}\,\hat{U}_1\, | \FT{\alpha} \rangle / g_{jkl}^{(v\FT{,\alpha})} = a^{(3)}_4\,a^{(3)}_3\,a^{(3)}_2 \label{eq73}\\
f^{jkl}_{34|2}\equiv\langle \FT{\alpha} | \,\FT{\hat{U}_4}\, \hat{a}\,\hat{U}_3\,\hat{a}\,\hat{U}_2\,\hat{a}^\dagger\,\hat{U}_1\, | \FT{\alpha} \rangle / g_{jkl}^{(v\FT{,\alpha})} = \nonumber\\ \chi_2\,a^{(3)}_4+\chi_{23}\,a^{(3)}_3+a^{(3)}_4\,a^{(3)}_3\,c^{(3)}_2 \label{eq74}\\
f^{jkl}_{4|23}\equiv\langle \FT{\alpha} | \,\FT{\hat{U}_4}\, \hat{a}\, \hat{U}_3\,\hat{a}^\dagger\,\hat{U}_2\,\hat{a}^\dagger\,\hat{U}_1\, | \FT{\alpha} \rangle / g_{jkl}^{(v\FT{,\alpha})} = \nonumber\\ \chi_{3}\,c^{(3)}_2+\chi_{23}\,c^{(3)}_3+a^{(3)}_4\,c^{(3)}_3\,c^{(3)}_2 \,.\label{eq75}
\end{gather}
In the general expression of the response function, the above vibrational terms are multiplied by the electronic contribution $C^{jkl}_{1110}\,g_{jkl}^{(e)}$.

The \FT{terms that are nonzero for $\alpha=0$} resulting from the presence of the annihilation and creation operators in all positions but the second one are given by:
\begin{gather}
f^{jkl}_{34|1}\equiv\langle \FT{\alpha} | \,\FT{\hat{U}_4}\, \hat{a}\,\hat{U}_3\,\hat{a}\,\hat{U}_2\,\hat{U}_1\,\hat{a}^\dagger\, | \FT{\alpha} \rangle / g_{jkl}^{(v\FT{,\alpha})} = \nonumber\\ \chi_{12}\,a^{(3)}_4+\chi_{13}\,a^{(3)}_3+a^{(3)}_4\,a^{(3)}_3\,c^{(3)}_1 \label{eq76}\\
f^{jkl}_{4|13}\equiv\langle \FT{\alpha} | \,\FT{\hat{U}_4}\, \hat{a}\, \hat{U}_3\,\hat{a}^\dagger\,\hat{U}_2\,\hat{U}_1\,\hat{a}^\dagger\, | \FT{\alpha} \rangle / g_{jkl}^{(v\FT{,\alpha})} = \nonumber\\ \chi_{3}\,c^{(3)}_1+\chi_{13}\,c^{(3)}_3+a^{(3)}_4\,c^{(3)}_3\,c^{(3)}_1 \,.\label{eq77}
\end{gather}
In the general expression of the response function, the above vibrational terms are multiplied by the electronic contribution $C^{jkl}_{1101}\,g_{jkl}^{(e)}$.

The \FT{terms that are nonzero for $\alpha=0$} resulting from the presence of the annihilation and creation operators are in all positions but the third one are given by:
\begin{gather}
f^{jkl}_{24|1}\equiv\langle \FT{\alpha} | \,\FT{\hat{U}_4}\, \hat{a}\,\hat{U}_3\,\hat{U}_2\,\hat{a}\,\hat{U}_1\,\hat{a}^\dagger\, | \FT{\alpha} \rangle / g_{jkl}^{(v\FT{,\alpha})} = \nonumber\\ \chi_1\,a^{(3)}_4+\chi_{13}\,a^{(3)}_2+a^{(3)}_4\,a^{(3)}_2\,c^{(3)}_1 \label{eq78}\\
f^{jkl}_{4|12}\equiv\langle \FT{\alpha} | \,\FT{\hat{U}_4}\, \hat{a}\, \hat{U}_3\,\hat{U}_2\,\hat{a}^\dagger\,\hat{U}_1\,\hat{a}^\dagger\, | \FT{\alpha} \rangle / g_{jkl}^{(v\FT{,\alpha})} = \nonumber\\ \chi_{23}\,c^{(3)}_1+\chi_{13}\,c^{(3)}_2+a^{(3)}_4\,c^{(3)}_2\,c^{(3)}_1\,.\label{eq79}
\end{gather}
In the general expression of the response function, the above vibrational terms are multiplied by the electronic contribution $C^{jkl}_{1011}\,g_{jkl}^{(e)}$.

The \FT{terms that are nonzero for $\alpha=0$} resulting from the presence of the annihilation and creation operators are in all positions but the fourth one are given by:
\begin{gather}
f^{jkl}_{|123}\equiv\langle \FT{\alpha} | \,\FT{\hat{U}_4}\, \hat{U}_3\,\hat{a}^\dagger\,\hat{U}_2\,\hat{a}^\dagger\,\hat{U}_1\,\hat{a}^\dagger\, | \FT{\alpha} \rangle / g_{jkl}^{(v\FT{,\alpha})} = c^{(3)}_3c^{(3)}_2c^{(3)}_1 \label{eq80}\\
f^{jkl}_{23|1}\equiv\langle \FT{\alpha} | \,\FT{\hat{U}_4}\,\hat{U}_3\,\hat{a}\,\hat{U}_2\,\hat{a}\,\hat{U}_1\,\hat{a}^\dagger\, | \FT{\alpha} \rangle / g_{jkl}^{(v\FT{,\alpha})} = \nonumber\\ \chi_1\,a^{(3)}_3+\chi_{12}\,a^{(3)}_2+a^{(3)}_3\,a^{(3)}_2\,c^{(3)}_1 \label{eq81}\\
f^{jkl}_{2|13}\equiv\langle \FT{\alpha} | \,\FT{\hat{U}_4}\, \hat{U}_3\,\hat{a}^\dagger\,\hat{U}_2\,\hat{a}\,\hat{U}_1\,\hat{a}^\dagger\, | \FT{\alpha} \rangle / g_{jkl}^{(v\FT{,\alpha})} = \nonumber\\ \chi_{2}\,c^{(3)}_1+c^{(3)}_3\,a^{(3)}_2\,c^{(3)}_1\label{eq81p}\\
f^{jkl}_{3|12}\equiv\langle \FT{\alpha} | \,\FT{\hat{U}_4}\, \hat{U}_3\,\hat{a}\,\hat{U}_2\,\hat{a}^\dagger\,\hat{U}_1\,\hat{a}^\dagger\, | \FT{\alpha} \rangle / g_{jkl}^{(v\FT{,\alpha})} = \nonumber\\ \chi_{2}\,c^{(3)}_1+\chi_{12}\,c^{(3)}_2+a^{(3)}_3\,c^{(3)}_2\,c^{(3)}_1\, .\label{eq82}
\end{gather}
In the general expression of the response function, the above vibrational terms are multiplied by the electronic contribution $C^{jkl}_{0111}\,g_{jkl}^{(e)}$

\subsection{{\color{black}Quartic} HT terms}
The HT terms are of order 0 and 4 in $\hat{\mu}_0$ and $\hat{\mu}_1$, respectively. The \FT{terms that are nonzero for $\alpha=0$}, resulting from the presence of the annihilation and creation operators in the four different positions, are given by: 
\begin{gather}
f^{jkl}_{234|1}\!\equiv\!\langle \FT{\alpha} | \,\FT{\hat{U}_4}\, \hat{a}\,\hat{U}_3\,\hat{a}\,\hat{U}_2\,\hat{a}\,\hat{U}_1\, \hat{a}^\dagger\,| \FT{\alpha} \rangle / g_{jkl}^{(v\FT{,\alpha})} \!=\! a^{(3)}_4 a^{(3)}_3 a^{(3)}_2 c^{(3)}_1\nonumber\\ +\chi_{1}\,a^{(3)}_4 a^{(3)}_3 +\chi_{12}\,a^{(3)}_4 a^{(3)}_2+\chi_{13}\, a^{(3)}_3 a^{(3)}_2  \label{eq83}\\
f^{jkl}_{34|12}\!\equiv\!\langle \FT{\alpha} | \,\FT{\hat{U}_4}\, \hat{a}\,\hat{U}_3\,\hat{a}\,\hat{U}_2\,\hat{a}^\dagger\,\hat{U}_1\, \hat{a}^\dagger\,| \FT{\alpha} \rangle / g_{jkl}^{(v\FT{,\alpha})} \!=\! 
 a^{(3)}_4 a^{(3)}_3 c^{(3)}_2 c^{(3)}_1 \nonumber\\
+ \chi_2 \left[ a^{(3)}_4 c^{(3)}_1 + \chi_{13}\right]   
+ \chi_{23} \left[a^{(3)}_3 c^{(3)}_1 + \chi_{12}\right]
\nonumber\\
+ \chi_{12} a^{(3)}_4 c^{(3)}_2
+ \chi_{13} a^{(3)}_3 c^{(3)}_2 \label{eq84}\\
f^{jkl}_{24|13}\!\equiv\!\langle \FT{\alpha} | \,\FT{\hat{U}_4}\, \hat{a}\,\hat{U}_3\,\hat{a}^\dagger\,\hat{U}_2\,\hat{a}\,\hat{U}_1\, \hat{a}^\dagger\,| \FT{\alpha} \rangle / g_{jkl}^{(v\FT{,\alpha})}
\!=\! a^{(3)}_4 c^{(3)}_3 a^{(3)}_2 c^{(3)}_1 \nonumber\\ +
\chi_3 \left[a^{(3)}_2 c^{(3)}_1 + \chi_1\right] 
+ \chi_{13} c^{(3)}_3 a^{(3)}_2  
+ \chi_1 a^{(3)}_4 c^{(3)}_3 \label{eq85}
\\
f^{jkl}_{4|123}\!\equiv\!\langle \FT{\alpha} | \,\FT{\hat{U}_4}\, \hat{a}\,\hat{U}_3\,\hat{a}^\dagger\,\hat{U}_2\,\hat{a}^\dagger\,\hat{U}_1\, \hat{a}^\dagger\,| \FT{\alpha} \rangle / g_{jkl}^{(v\FT{,\alpha})} 
\!=\! a^{(3)}_4 c^{(3)}_3 c^{(3)}_2 c^{(3)}_1 \nonumber\\ +\chi_{3}\,c^{(3)}_2 c^{(3)}_1+\chi_{23}\,c^{(3)}_3 c^{(3)}_1+\chi_{13}\,c^{(3)}_3 c^{(3)}_2\,. \label{eq86}
\end{gather}
In the general expression of the response function, the above vibrational terms are multiplied by the electronic contribution $C^{jkl}_{1111}\,g_{jkl}^{(e)}$.

\subsection{Relation between terms corresponding to different Feynman diagrams}

The propagator corresponding to the double-sided Feynman diagram where all the light-matter interactions take place on the left-hand side [such as the one in Fig. \ref{fig2}(b)] is:
\begin{gather}
    {\rm Tr}[ \hat{\mu}(t_{13})\,\hat{\mu}(t_{12})\,\hat{\mu}(t_1)\,\hat{\mu}(0)\hat{\rho}(0)] = \nonumber \\
    \langle\psi_0| \,\FT{\hat{\mathcal{U}}^\dagger(t_{13})}\,\hat{\mu}\,\hat{\mathcal{U}}(t_3)\,\hat{\mu}\,\hat{\mathcal{U}}(t_2)\,\hat{\mu}\,\hat{\mathcal{U}}(t_1)\,\mu\,|\psi_0\rangle\,,
\end{gather}
being $\hat{\mu}=\hat{\mu}(0)$, the initial state $\hat{\rho}(0)=|\psi_0\rangle\langle\psi_0|$, \FT{$t_{12}\equiv t_1+t_2$, and $t_{13}\equiv t_1+t_2+t_3$}.

The propagator corresponding to the double-sided Feynman diagram where the first and fourth light-matter interactions take place on the left-hand side, and the second and third interactions take place on the right [such as the one in Fig. \ref{fig2}(c)] is:
\begin{gather}
    {\rm Tr}[\hat{\mu}(t_{13})\,\hat{\mu}(0)\,\hat{\rho}(0)\,\hat{\mu}(t_1)\,\hat{\mu}(t_{12})] = \nonumber\\
    \langle \psi_0|\,\FT{\hat{\mathcal{U}}^\dagger(t_{1})}\,\hat{\mu}\,\hat{\mathcal{U}}^\dagger(t_2)\,\hat{\mu}\,\hat{\mathcal{U}}^\dagger (t_3)\,\hat{\mu}\,\hat{\mathcal{U}}(t_{13})\,\hat{\mu}\,|\psi_0\rangle \,,
\end{gather}
where we have exploited the invariance of the trace with respect to cyclic permutations.
The second propagator can be obtained from the first one simply by the transformation: $t_1\rightarrow t_1+t_2+t_3$, $t_2\rightarrow -t_3$, $t_3\rightarrow -t_2$. \FT{We note that this correspondence applies to all initial states $|\psi_0\rangle$, and thus also to the case where the vibrational mode is initialized in a generic coherent state $|\alpha\rangle$.}

{\color{black}
\subsection{Simulated 2D spectra}
Here we derive the contributions to the third order ($M=3$) response functions of a system with three electronic levels that are of lowest nonzero order in the displacement parameters $z_k$ ($k=1,2$, being $z_0=0$) and $\hat{\mu}_1$. These can be grouped in three terms. The first one arises from the linear HT terms that are of first order also in $z_k$ (these parameters appear in the functions $f_{k,j}$, with $j=1,2,3$). The second term is proportional to $z_k^2$ or $z_1 z_2$, and is of the zero-th order in $\hat{\mu}_1$ (FC contributions). The third term arises from the quadratic HT terms, and is of the zero-th order in the displacements $z_k$. The vibrational mode is assumed to be initialized in the ground state ($\alpha=0$). All the above contributions are expressed hereafter in terms of the oscillating functions $\chi_k$ and $\chi_{kl}$, defined in Eq. (\ref{eq12}).}

{\color{black}The plotted response functions are obtained by setting to zero one of the waiting times (either $t_1$ or $t_2$, resulting in $\chi_1=1$ and $\chi_2=1$, respectively) and by performing a Fourier transform with respect to the other two. As a result of the Fourier transform, one has that\cite{Hamm2011}: 
\begin{align}\label{eqhhh}
e^{\pm i\omega_et_k}\chi_k &\longrightarrow \frac{1}{\gamma - i(\omega_k\pm\omega_e-\omega)}\nonumber\\\ e^{\pm i\omega_et_k}\chi_k^* &\longrightarrow \frac{1}{\gamma - i(\omega_k\pm\omega_e+\omega)}\,,
\end{align}
where $k=1,2,3$, $\hbar\omega$ is the vibrational energy, and $\gamma$ is a phenomenological constant that accounts for the Lorentzian line broadening. The spectra that are plotted in Section \ref{s4} are obtained by performing the above replacements on the $\chi_k$ that are subject to Fourier transform in the function $g^{(e)}_{jkl}\,g^{(v)}_{jkl}\sum_{p=0}^4 r_{jkl,p}\approx g^{(e)}_{jkl}(\tilde A + \tilde B + \tilde C)$ .}

{\color{black}
\subsubsection{Ground state bleaching}
We start by considering the rephasing term of the ground state bleaching response function for a system with three electronic levels [diagram in Fig. \ref{fig3} (a)]. This includes contributions that are linear in $\hat{\mu}_1$ and in $z_1$, coming from $r_{101,1}$ [Eq. \ref{eqx01}] and from the first term in the Taylor expansion of $g^{(v)}_{101} $ [corresponding to 1, see Eq. \ref{eq91}]:
\begin{gather}
\tilde{A}_1 \!=\! z_1\!\left[ C^{101}_{0001} \left(\chi_2^* \!-\! \chi_{12}^*  \!+\! 1 \!-\! \chi_3\right) 
\!+\! C^{101}_{0010} \left(\chi_{23}^* \!-\! \chi_{13}^* \!-\! 1 \!+\! \chi_{3}\right) \right.\nonumber\\
+ C^{101}_{0100} \left(1\!-\! \chi_{1}^* \!+\! \chi_{2}^* \!-\! \chi_{23}^*\right)
\left.+ C^{101}_{1000} \left( \chi_{1}^* \!-\! 1 \!+\! \chi_{12}^* \!-\! \chi_{13}^* \right) \right],\label{eqA1}
\end{gather}
where $C^{101}_{1000}=C^{101}_{0010}$ and $C^{101}_{0001}=C^{101}_{0100}$.}

{\color{black}The contributions that are quadratic in $z_1$ come from $r_{101,0}$ [Eq. \ref{eqy01}] and from the second term in the Taylor expansion of $g^{(v)}_{101}=e^f$, corresponding to:
\begin{gather}
     \tilde{B}_1 \!=\! C^{101}_{0000} f = C^{101}_{0000} z_1^2 \left(\chi_1^* \!+\! \chi_3 \!-\!\chi_2^* \!+\! \chi_{23}^* \!+\! \chi_{12}^*\!-\!\chi_{13}^*\!-\!2\right).\label{eqB1}
\end{gather}}

{\color{black}Finally, the contribution that is quadratic in $\hat{\mu}_1$ and of zero order in $z_1$ come from $r_{101,2}$ [Eq. \ref{eqz01}] and from the first term in the Taylor expansion of $g^{(v)}_{101} $ [corresponding to 1, see Eq. \ref{eq91}]:
\begin{gather}
\tilde{C}_1 = C^{101}_{0011}\chi_3 + C^{101}_{0101}\chi_{2}^* + C^{101}_{1001}\chi_{12}^* \nonumber\\+ C^{101}_{0110}\chi_{23}^* + C^{101}_{1010}\chi_{13}^* + C^{101}_{1100}\chi_1^*,\label{eqC1}
\end{gather}
where $C^{101}_{0011}\!=\!C^{101}_{1100}\!=\!C^{101}_{1001}$ and $C^{101}_{0110}\!=\!C^{101}_{1001}\!=\!C^{101}_{1100}$.
The three equations displayed above are derived from the ones reported in Subsec. \ref{subsec:torf} and Appendix \ref{ggg} by applying the following replacements: $\chi_1 \rightarrow \chi_3$, $\chi_2 \rightarrow \chi_{23}^*$, $\chi_3 \rightarrow \chi^*_1$, $\chi_{12}\rightarrow\chi_{2}^*$, $\chi_{23}\rightarrow\chi_{13}^*$, and $\chi_{13}\rightarrow\chi_{12}^*$.}

{\color{black}We next considering the non-rephasing term of the ground state bleaching response function [diagram in Fig. \ref{fig3} (b)]. This includes contributions that are linear in $\hat{\mu}_1$ and in $z_1$, coming from $r_{101,1}$ [Eq. \ref{eqx01}] and from the first term in the Taylor expansion of $g^{(v)}_{101} $ [corresponding to 1, see Eq. \ref{eq91}]:
\begin{gather}
\tilde{A}_2 \!=\! z_1\! \left[ C^{101}_{0001} \left(1\!-\!\chi_{1} \!-\! \chi_{13} \!+\! \chi_{12} \right) 
\!+\!C^{101}_{0010} \left( \chi_{2} \!-\! \chi_{23} \!-\! 1 \!+\! \chi_{1}\right)\right.\nonumber\\ \left.
+C^{101}_{0100} \left(\chi_{12} \!+\! 1 \!-\! \chi_3 \!-\! \chi_{2}\right)
+C^{101}_{1000} \left(\chi_{13} \!-\! \chi_{23} \!+\! \chi_{3} \!-\! 1 \right) \right].\label{eqA2}
\end{gather}}

{\color{black}The contributions that are quadratic in $z_1$ come from $r_{101,0}$ [Eq. \ref{eqy01}] and from the second term in the Taylor expansion of $g^{(v)}_{101} =e^f$, corresponding to:
\begin{gather}
     \tilde{B}_2 = C^{101}_{0000} f = C^{101}_{0000} z_1^2 \left(\chi_1\! +\! \chi_3\! +\!\chi_2 \!-\! \chi_{23}\!-\!\chi_{12}\! +\! \chi_{13}\!-\!2\right).\label{eqB2}
\end{gather}}

{\color{black}Finally, the contribution that is quadratic in $\hat{\mu}_1$ and of zero order in $z_1$ come from $r_{101,2}$ [Eq. \ref{eqz01}] and from the first term in the Taylor expansion of $g^{(v)}_{101} $ [Eq. \ref{eq91}]:
\begin{gather}
\tilde{C}_2 = C^{101}_{0011}\chi_1 + C^{101}_{0101}\chi_{12} + C^{101}_{1001}\chi_{13} \nonumber\\+ C^{101}_{0110}\chi_{2} + C^{101}_{1010}\chi_{23} + C^{101}_{1100}\chi_3.\label{eqC2}
\end{gather}
The three equations displayed above are derived from the ones reported in Subsec. \ref{subsec:torf} and Appendix \ref{ggg} without applying any replacement in the waiting times.}

{\color{black}
\subsubsection{Stimulated emission}
The rephasing term of the stimulated emission response function [diagram in Fig. \ref{fig3} (c)] includes contributions that are linear in $\hat{\mu}_1$ and in $z_1$, coming from $r_{101,1}$ [Eq. \ref{eqx01}] and from the first term in the Taylor expansion of $g^{(v)}_{101} $ [corresponding to 1, see Eq. \ref{eq91}]:
\begin{gather}
\tilde{A}_3\!=\!z_{1} \left[C^{101}_{0001} \left(1\!-\!\chi_{1}^*\! -\! \chi_{23}\! +\!  \chi_{2}\right)
\!+\! C^{101}_{0010} \left(\chi_{3}^* \!-\! \chi_{13}^* \!-\! 1 \!+\! \chi_{23}\right) \right.\nonumber\\
\left. + C^{101}_{0100} \left(1 \!-\! \chi_{3}^* \!-\! \chi_{12}^* \!+\! \chi_{2}\right)
+ C^{101}_{1000} \left(\chi_{1}^* \!+\! \chi_{12}^* \!-\! \chi_{13}^* \!-\! 1\right)\right].\label{eqA3}
\end{gather}}

{\color{black}The contributions that are quadratic in $z_1$ come from $r_{101,0}$ [Eq. \ref{eqy01}] and from the second term in the Taylor expansion of $g^{(v)}_{101} =e^f$, corresponding to :
\begin{gather}
\tilde{B}_3\!=\!C^{101}_{0000} f \!=\! C^{101}_{0000} z_1^2 \left( \chi_{12}^*\! +\! \chi_{23}\! +\! \chi_3^* \!-\! \chi_2 \!+\! \chi_1^* \!-\! \chi_{13}^*\!-\!2\right).\label{eqB3}
\end{gather}

Finally, the contribution that is quadratic in $\hat{\mu}_1$ and of zero order in $z_1$ come from $r_{101,2}$ [Eq. \ref{eqz01}] and from the first term in the Taylor expansion of $g^{(v)}_{101} $ [Eq. \ref{eq91}]:
\begin{gather}
\tilde{C}_3=C^{101}_{0011}\chi_{23} + C^{101}_{0101}\chi_{2} + C^{101}_{1001}\chi_{1}^* \nonumber\\+ C^{101}_{0110}\chi_{3}^* + C^{101}_{1010}\chi_{13}^* + C^{101}_{1100}\chi_{12}^*.\label{eqC3}
\end{gather}
The three equations displayed above are derived from the ones reported in Subsec. \ref{subsec:torf} and Appendix \ref{ggg} by applying the following replacements: $\chi_1 \rightarrow \chi_{23}$, $\chi_2 \rightarrow \chi^*_3$, $\chi_3 \rightarrow \chi_{12}^*$, $\chi_{12}\rightarrow\chi_{2}$, $\chi_{23}\rightarrow\chi_{13}^*$, and $\chi_{13}\rightarrow\chi_{1}^*$.}

{\color{black}
We next considering the non-rephasing term of the stimulated emission response function [diagram in Fig. \ref{fig3} (d)]. This includes contributions that are linear in $\hat{\mu}_1$ and in $z_1$, coming from $r_{101,1}$ [Eq. \ref{eqx01}] and from the first term in the Taylor expansion of $g^{(v)}_{101} $ [corresponding  to 1, see Eq. \ref{eq91}]:
\begin{gather}
\tilde{A}_4\!=\!z_{1}\!\left[C^{101}_{0001}\left(\chi_{12} \!-\! \chi_{1} \! +\! 1 \!-\! \chi_{13}\right) 
\!+\! C^{101}_{0010} \left(\chi_{3}^* \!-\! \chi_{23}^* \!-\! 1 \!+\! \chi_{13}\right) \right.\nonumber\\
\left. + C^{101}_{0100} \left(1\!-\! \chi_{2}^* \!-\! \chi_{3}^* \!+\! \chi_{12} \right) 
+ C^{101}_{1000} \left(\chi_{2}^* \!-\! \chi_{23}^* \!- 1 \!+\! \chi_{1}\right) \right].\label{eqA4}
\end{gather}}

{\color{black}The contributions that are quadratic in $z_1$ come from $r_{101,0}$ [Eq. \ref{eqy01}] and from the second term in the Taylor expansion of $g^{(v)}_{101} =e^f$, corresponding to:
\begin{gather}
\tilde{B}_4\!=\!C^{101}_{0000}  f \!=\! C^{101}_{0000} z_1^2 C^{101}_{0000} \nonumber\\ \left(\chi_{13} \!+\! \chi_2^* \!+\! \chi_3^* \!+\! \chi_1 \!-\! \chi_{23}^* \!-\! \chi_{12} \!-\! 2\right).\label{eqB4}
\end{gather}

Finally, the contribution that is quadratic in $\hat{\mu}_1$ and of zero order in $z_1$ come from $r_{101,2}$ [Eq. \ref{eqz01}] and from the first term in the Taylor expansion of $g^{(v)}_{101} $ [Eq. \ref{eq91}]:
\begin{gather}
\tilde{C}_4=C^{101}_{0011}\chi_{13} + C^{101}_{0101}\chi_{12} + C^{101}_{1001}\chi_{1} \nonumber\\+ C^{101}_{0110}\chi_{3}^* + C^{101}_{1010}\chi^*_{23} + C^{101}_{1100}\chi_2^*.\label{eqC4}
\end{gather}
The three equations displayed above are derived from the ones reported in Subsec. \ref{subsec:torf} and Appendix \ref{ggg} by applying the following replacements: $\chi_1 \rightarrow \chi_{13}$, $\chi_2 \rightarrow \chi_3^*$, $\chi_3 \rightarrow \chi_2^*$, $\chi_{12}\rightarrow\chi_{12}$, $\chi_{23}\rightarrow\chi_{23}^*$, and $\chi_{13}\rightarrow\chi_{1}$.}

{\color{black}
\subsubsection{Excited state absorption}
The rephasing term of the excited state absorption response function [diagram in Fig. \ref{fig3} (e)] includes contributions that are linear in $\hat{\mu}_1$ and in $z_1$, coming from $r_{121,1}$ [Eq. \ref{eqx01}] and from the first term in the Taylor expansion of $g^{(v)}_{121} $ [corresponding to 1, see Eq. \ref{eq91}]:
\begin{gather}
\tilde{A}_5=C^{121}_{0001} \left[ z_1\left(1\!-\! \chi_{2} \!-\! \chi_{1}^* \!+\! \chi_{23}\right) + z_{2} \left(\chi_{2} \!-\! \chi_{23}\right) \right]\nonumber\\
+ C^{121}_{0010}\left[ z_1\left(\chi_{3} \!+\! \chi_{2} \!-\! \chi_{12}^* \!-\! 1\right) + z_{2} \left(1\!-\! \chi_{3}\right)\right]\nonumber\\
+ C^{121}_{0100}\left[ z_{1} \left(\chi_{23} \!-\! \chi_{13}^* \!+\! 1 \!-\! \chi_{3}\right) + z_{2} \left(\chi_{3}\!-\!1\right)\right]\nonumber\\
+ C^{121}_{1000}\left[ z_{1} \left(\chi_{13}^* \!-\! 1 \!+\! \chi_{1}^* \!-\! \chi_{12}^*\right)  + z_{2} \left(\chi_{12}^* \!-\! \chi_{13}^*\right)  \right].\label{eqA5}
\end{gather}
The contributions that are quadratic in $z_1$ come from $r_{121,0}$ [Eq. \ref{eqy01}] and from the second term in the Taylor expansion of $g^{(v)}_{121} =e^f$, corresponding to:
\begin{gather}
\tilde{B}_5\!=\!C^{121}_{0000} f = C^{121}_{0000}\left[z_{12}^2 (\chi_3\!-\!1) + z_1 z_{12}  \right.\nonumber\\(\chi_2\! -\! \chi_{23} \!-\! \chi_{12}^* \!+\! \chi_{13}^*)  \left.+ z_1^2 (\chi_1^*\!-\!1)\right].\label{eqB5}
\end{gather}
Finally, the contribution that is quadratic in $\hat{\mu}_1$ and of zero order in $z_1$ come from $r_{121,2}$ [Eq. \ref{eqz01}] and from the first term in the Taylor expansion of $g^{(v)}_{121} $ [Eq. \ref{eq91}]:
\begin{gather}
\tilde{C}_5=C^{121}_{0011}\chi_2 + C^{121}_{0101}\chi_{23} + C^{121}_{1001}\chi_{1}^* \nonumber\\+ C^{121}_{0110}\chi_{3} + C^{121}_{1010}\chi_{12}^* + C^{121}_{1100}\chi_{13}^*.\label{eqC5}
\end{gather}
The three equations displayed above are derived from the ones reported in Subsec. \ref{subsec:torf} and Appendix \ref{ggg} by applying the following replacements: $\chi_1 \rightarrow \chi_2$, $\chi_2 \rightarrow \chi_3$, $\chi_3 \rightarrow \chi_{13}^*$, $\chi_{12}\rightarrow\chi_{23}$, $\chi_{23}\rightarrow\chi_{12}^*$, and $\chi_{13}\rightarrow\chi_{1}^*$.}

{\color{black}The non-rephasing term of the excited state absorption response function [diagram in Fig. \ref{fig3} (f)] includes contributions that are linear in $\hat{\mu}_1$ and in $z_1$, coming from $r_{121,1}$ [Eq. \ref{eqx01}] and from the first term in the Taylor expansion of $g^{(v)}_{121} $ [corresponding to 1, see Eq. \ref{eq91}]:
\begin{gather}
\tilde{A}_6=C^{121}_{0001}\left[  z_{1}\left(1\!-\!\chi_{12}\! -\! \chi_{1}\! +\! \chi_{13}\right)+ z_{2} \left( \chi_{12} \!-\! \chi_{13}\right)\right] \nonumber\\
+C^{121}_{0010} \left[z_{1}\left(\chi_{12}\!-\! \chi_{2}^* \!-\! 1 \!+\! \chi_{3}\right) + z_{2}\left(1 \!-\! \chi_{3}\right)\right]\nonumber\\
+ C^{121}_{0100} \left[z_{1}\left(\chi_{13}\! -\! \chi_{23}^* \!+\! 1 \!-\! \chi_{3}\right) + z_{2} \left(\chi_{3}\!-\!1\right)\right]\nonumber\\
+ C^{121}_{1000} \left[z_{1}\left(\chi_{23}^* \!-\! \chi_{2}^* \!-\! 1 \!+\! \chi_{1}\right) + z_{2}\left( \chi_{2}^* \!-\! \chi_{23}^*\right)\right].\label{eqA6}
\end{gather}
The contributions that are quadratic in $z_1$ come from $r_{121,0}$ [Eq. \ref{eqy01}] and from the second term in the Taylor expansion of $g^{(v)}_{121} =e^f$, corresponding to:
\begin{gather}
\tilde{B}_6=C^{121}_{0000}f=C^{121}_{0000}\left[z_{21}^2 (\chi_3\!-\!1) + z_1 z_{21} \right.\nonumber\\(\chi_2^* \!-\! \chi_{23}^* \!-\! \chi_{12} \!+\! \chi_{13})  \left.+ z_1^2 (\chi_1\!-\!1)\right].\label{eqB6}
\end{gather}
Finally, the contribution that is quadratic in $\hat{\mu}_1$ and of zero order in $z_1$ come from $r_{121,2}$ [Eq. \ref{eqz01}] and from the first term in the Taylor expansion of $g^{(v)}_{121} $ [Eq. \ref{eq91}]:
\begin{gather}
\tilde{C}_6=C^{121}_{0011}\chi_{12} + C^{121}_{0101}\chi_{13} + C^{121}_{1001}\chi_{1} \nonumber\\+ C^{121}_{0110}\chi_{3} + C^{121}_{1010}\chi_{2}^* + C^{121}_{1100}\chi_{23}^*.\label{eqC6}
\end{gather}
The three equations displayed above are derived from the ones reported in Subsec. \ref{subsec:torf} and Appendix \ref{ggg} by applying the following replacements: $\chi_1 \rightarrow \chi_{12}$, $\chi_2 \rightarrow \chi_3$, $\chi_3 \rightarrow \chi_{23}^*$, $\chi_{12}\rightarrow\chi_{13}$, $\chi_{23}\rightarrow\chi_{2}^*$, and $\chi_{13}\rightarrow\chi_{1}$.}

{\color{black}\subsubsection{Double quantum coherence}
The first term of the double quantum coherence response function [diagram in Fig. \ref{fig3} (g)] includes contributions that are linear in $\hat{\mu}_1$ and in $z_1$, coming from $r_{121,1}$ [Eq. \ref{eqx01}] and from the first term in the Taylor expansion of $g^{(v)}_{121} $ [corresponding to 1, see Eq. \ref{eq91}]:
\begin{gather}
\tilde{A}_7=C^{121}_{0001} \left[z_{1} \left(1\!-\!\chi_{1}\! +\! \chi_{13}\! -\! \chi_{12}\right) + z_{2} \left(\chi_{1}\! -\! \chi_{13} \right)\right] \nonumber\\
+ C^{121}_{0010} \left[ z_{1} \left(\chi_{23}\! -\! 1\! -\! \chi_{2} \!+\! \chi_{1}\right) + z_{2} \left(1\!-\!\chi_{23}\right)\right]\nonumber\\
+ C^{121}_{0100} \left[z_{1}\left(1\!-\! \chi_{3}^* \!-\! \chi_{23} \!+\!\chi_{13}\right) + z_{2} \left(\chi_{23} \!-\! 1\right)\right]\nonumber\\
+ C^{121}_{1000} \left[ z_{1} \left( \chi_{3}^* \!+\! \chi_{12} \!-\! 1 \!-\! \chi_{2}\right) +z_2\left(\chi_{2} \!-\! \chi_{3}^* \right)\right].\label{eqA7}
\end{gather}
The contributions that are quadratic in $z_1$ come from $r_{121,0}$ [Eq. \ref{eqy01}] and from the second term in the Taylor expansion of $g^{(v)}_{121} =e^f$, corresponding to:
\begin{gather}
\tilde{B}_7=C^{121}_{0000}f=C^{121}_{0000}\left[z_1 z_{12} \right.\nonumber\\(\chi_3^*\! -\! \chi_2\! +\! \chi_1\! -\!\chi_{13})+ z_{21}^2 (\chi_{23}\!-\!1)  \left.+ z_1^2 (\chi_{12}\!-\!1) \right].\label{eqB7}
\end{gather}
Finally, the contribution that is quadratic in $\hat{\mu}_1$ and of zero order in $z_1$ come from $r_{121,2}$ [Eq. \ref{eqz01}] and from the first term in the Taylor expansion of $g^{(v)}_{121} $ [Eq. \ref{eq91}]:
\begin{gather}
\tilde{C}_7=C^{121}_{0011}\chi_1 + C^{121}_{0101}\chi_{13} + C^{121}_{1001}\chi_{12} \nonumber\\+ C^{121}_{0110}\chi_{23} + C^{121}_{1010}\chi_{2} + C^{121}_{1100}\chi_3^*.\label{eqC7}
\end{gather}
The three equations displayed above are derived from the ones reported in Subsec. \ref{subsec:torf} and Appendix \ref{ggg} by applying the following replacements: $\chi_1 \rightarrow \chi_1$, $\chi_2 \rightarrow \chi_{23}$, $\chi_3 \rightarrow \chi_3^*$, $\chi_{12}\rightarrow\chi_{13}$, $\chi_{23}\rightarrow\chi_{2}$, and $\chi_{13}\rightarrow\chi_{12}$.}

{\color{black}The second term of the double quantum coherence response function [diagram in Fig. \ref{fig3} (h)] includes contributions that are linear in $\hat{\mu}_1$ and in $z_1$, coming from $r_{121,1}$ [Eq. \ref{eqx01}] and from the first term in the Taylor expansion of $g^{(v)}_{121} $ [corresponding to 1, see Eq. \ref{eq91}]:
\begin{gather}
\tilde{A}_8=C^{121}_{0001} \left[z_{1} \left(\chi_{12}\! -\! \chi_{13}\! +\! 1 \!-\! \chi_{1}\right) + z_{2}\left(\chi_{1}\! -\! \chi_{12} \right)\right] \nonumber\\
+ C^{121}_{0010} \left[ z_{1} \left(\chi_{2}\! +\! \chi_{1}\! -\! \chi_{23}\! -\! 1\right) + z_{2}\left(1 \!-\! \chi_{2}\right)\right]\nonumber\\
+C^{121}_{0100} \left[ z_{1} \left(\chi_{12}\! +\! 1\! -\! \chi_{3}\! -\! \chi_{2}\right) +z_2\left(\chi_{2}\!-\!1\right)\right]\nonumber\\
+ C^{121}_{1000} \left[ z_{1}\left(\chi_{13}\! -\! \chi_{23}\! -\! 1 \!+\! \chi_{3} \right) + z_{2}\left( \chi_{23}\! -\! \chi_{3} \right)\right].\label{eqA8}
\end{gather}
The contributions that are quadratic in $z_1$ come from $r_{121,0}$ [Eq. \ref{eqy01}] and from the second term in the Taylor expansion of $g^{(v)}_{121} =e^f$, corresponding to:
\begin{gather}
\tilde{B}_8=C^{111}_{0000} f = C^{111}_{0000}\left[z_{21}^2 (\chi_2\!-\!1) + \right.\nonumber\\z_1 z_{12} (\chi_3 \!+\! \chi_1 \!-\! \chi_{23}\! -\! \chi_{12})  \left.+ z_1^2(\chi_{13}\!-\!1)\right].\label{eqB8}
\end{gather}
Finally, the contribution that is quadratic in $\hat{\mu}_1$ and of zero order in $z_1$ come from $r_{121,2}$ [Eq. \ref{eqz01}] and from the first term in the Taylor expansion of $g^{(v)}_{121} $ [Eq. \ref{eq91}]:
\begin{gather}
\tilde{C}_8=C^{121}_{0011}\chi_1 + C^{121}_{0101}\chi_{12} + C^{121}_{1001}\chi_{13} \nonumber\\+ C^{121}_{0110}\chi_{2} + C^{121}_{1010}\chi_{23} + C^{121}_{1100}\chi_3.\label{eqC8}
\end{gather}
The three equations displayed above are derived from the ones reported in Subsec. \ref{subsec:torf} and Appendix \ref{ggg} without applying any replacement in the waiting times.}


%

\end{document}